\documentclass[preprint,prl,endfloats*,superscriptaddress]{revtex4-1}

\usepackage{graphicx}
\usepackage{dcolumn}
\usepackage{bm}
\usepackage{physics,amsmath}
\usepackage{latexsym}

\usepackage{color,soul} 

\newcommand{\rfig}[1]{Fig.~\ref{#1}}

\makeindex

\begin{document}
	
	\title{Antisymmetric Seebeck Effect in a Tilted Weyl Semimetal}
	
	\author{Bingyan Jiang}
	\author{Jiaji Zhao}
	\affiliation{State Key Laboratory for Artificial Microstructure and Mesoscopic Physics, Frontiers Science Center for Nano-optoelectronics, Peking University, Beijing 100871, China}
	\author{Jiangyuan Qian}
	\affiliation{Shenzhen Institute for Quantum Science and Engineering and Department of Physics,
		Southern University of Science and Technology (SUSTech), Shenzhen 518055, China}
	\affiliation{Shenzhen Key Laboratory of Quantum Science and Engineering, Shenzhen 518055, China}
	\author{Shen Zhang}
	\affiliation{Beijing National Laboratory for Condensed Matter Physics, Institute of Physics, Chinese Academy of Sciences, Beijing 100190, China}
	\author{XiaoBin Qiang}
	\affiliation{Shenzhen Institute for Quantum Science and Engineering and Department of Physics,
		Southern University of Science and Technology (SUSTech), Shenzhen 518055, China}
	\affiliation{Shenzhen Key Laboratory of Quantum Science and Engineering, Shenzhen 518055, China}
	\author{Lujunyu Wang}
	\author{Ran Bi}
	\author{Juewen Fan}
	\affiliation{State Key Laboratory for Artificial Microstructure and Mesoscopic Physics, Frontiers Science Center for Nano-optoelectronics, Peking University, Beijing 100871, China}
	\author{Hai-Zhou Lu}
	\email{luhz@sustech.edu.cn}
	\affiliation{Shenzhen Institute for Quantum Science and Engineering and Department of Physics,
		Southern University of Science and Technology (SUSTech), Shenzhen 518055, China}
	\affiliation{Shenzhen Key Laboratory of Quantum Science and Engineering, Shenzhen 518055, China}
	\author{Enke Liu}
	\email{ekliu@iphy.ac.cn}
	\affiliation{Beijing National Laboratory for Condensed Matter Physics, Institute of Physics, Chinese Academy of Sciences, Beijing 100190, China}
	\affiliation{Songshan Lake Materials Laboratory, Dongguan 523808, China}
	\author{Xiaosong Wu}
	\email{xswu@pku.edu.cn}
	\affiliation{State Key Laboratory for Artificial Microstructure and Mesoscopic Physics, Frontiers Science Center for Nano-optoelectronics, Peking University, Beijing 100871, China}
	\affiliation{Collaborative Innovation Center of Quantum Matter, Beijing 100871, China}
	\affiliation{Peking University Yangtze Delta Institute of Optoelectronics, Nantong 226010, Jiangsu, China}
	
	\begin{abstract}
		Tilting the Weyl cone breaks the Lorentz invariance and enriches the Weyl physics. Here, we report the observation of a magnetic-field-antisymmetric Seebeck effect in a tilted Weyl semimetal, Co$_3$Sn$_2$S$_2$. Moreover, it is found that the Seebeck effect and the Nernst effect are antisymmetric in both the in-plane magnetic field and the magnetization. We attribute these exotic effects to the one-dimensional chiral anomaly and phase space correction due to the Berry curvature. The observation is further reproduced by a theoretical calculation, taking into account the orbital magnetization.
	\end{abstract}
	
	
	
	\maketitle
	
	Since Weyl fermions were first predicted in high-energy particle physics over 90 years ago, their experimental confirmation has remained elusive until they were recently discovered as emergent quasiparticles in low-energy condensed matter systems\cite{Wan2011,Weng2015Mar,Huang2015Jun,Lv2015Jul,Xu2015Aug,Lv2015Sep,Xu2015Sep}. The discovery has immediately led to intensive research activities\cite{Armitage2018,Nagaosa2020}. In sharp contrast to Weyl fermions in free space, Weyl quasiparticles in condensed matter systems are not constrained by the Lorentz invariance. Breaking of this symmetry can be attained simply by a tilt of the Weyl cone, which is not uncommon in real crystals with discrete symmetries. It was then realized that a tilt can give rise to many unusual effects that are impossible for Weyl fermions in free space\cite{Volovik2016Nov,Carbotte2016Oct,Isobe2016Mar,Weststrom2017Oct,Farajollahpour2019Jun,Ferreiros2019Feb,Ferreiros2017,Chan2017Jan,Konig2017Aug,Ma2017Sep,Golub2018Aug,Steiner2017Jul,Zyuzin2018Jan,Nguyen2018Jun,Jia2019Jan,Trescher2015Mar,Zyuzin2017,Sharma2017,Das2019,Ma2019a,Das2019a,vanderWurff2019Jul,Ahn2020Nov,Jiang2021Jun,Wawrzik2021Jul}. For instance, tilted Weyl semimetals can be used to simulate black holes\cite{Volovik2016Nov} and curved spacetime\cite{Weststrom2017Oct,Farajollahpour2019Jun,Ferreiros2019Feb}. It may allow a unique mixed axial-torsional anomaly to be detected\cite{Ferreiros2019Feb}. Tilts of the Weyl spectra enable generation of a photocurrent that is forbidden in an ideal two-dimensional Dirac spectrum\cite{Chan2017Jan,Konig2017Aug,Ma2017Sep,Golub2018Aug}. Tilts can also induce a flat band in disorder-driven non-Hermitian Weyl semimetals\cite{Zyuzin2018Jan}. Nguyen $et\ al$. proposed that tilts could induce drastically different transport phenomena and lead to easy tuning of valley filtering and beam splitting\cite{Nguyen2018Jun}. Unusual quantum transport effects have also been predicted in tilted Weyl semimetals\cite{Trescher2015Mar,Zyuzin2017,Sharma2017,Das2019,Ma2019a,Das2019a,vanderWurff2019Jul}. Despite a large body of theoretical work, experimental studies on tilted Weyl semimetals are relatively few.
	
	Most of these effects are closely related to the Berry curvature of the Bloch band. In electrical transport, the major effect of the Berry curvature, in the presence of an electric field, manifests as an anomalous velocity in electron dynamics. However, this mechanism is not applicable when the electric field is absent, for instance, in the case of thermoelectric transport. It was found that, in order to understand Berry-curvature-related thermoelectric effects, one has to consider the correction to orbital magnetization due to the Berry curvature\cite{Xiao2006}. As a result, thermoelectric measurements can be a distinctive tool for investigation of the Berry physics. Indeed, several unique effects in Weyl semimetals were uncovered in thermoelectric studies\cite{Jia2016,Gooth2017,Zhang2020}.
	
	Co$_3$Sn$_2$S$_2$ is a time reversal symmetry broken Weyl semimetal\cite{Liu2018,Liu2019,Morali2019}. It hosts three pairs of Weyl cones, which are linked by the $C_3$ symmetry. An earlier study has suggested that Weyl cones are tilted\cite{Jiang2021Jun}. One pair of cones are tilted along the $y$ axis. In this work, we study the thermoelectric effect of Co$_3$Sn$_2$S$_2$ and reveal a highly unusual thermoelectric transport phenomenon related to the tilt of the Weyl cone. In particular, the Seebeck coefficient displays a magnetic-field-antisymmetric contribution, resembling the symmetry of the Nernst effect. Moreover, the Seebeck coefficient and the Nernst signal are also antisymmetric in magnetization. It is shown that these phenomena originate from the one-dimensional chiral anomaly and phase space correction due to the Berry curvature, which is corroborated by our theoretical calculation including the orbital magnetization.
	
\begin{figure}[htbp]
		\begin{center}
			\includegraphics[width=1\columnwidth]{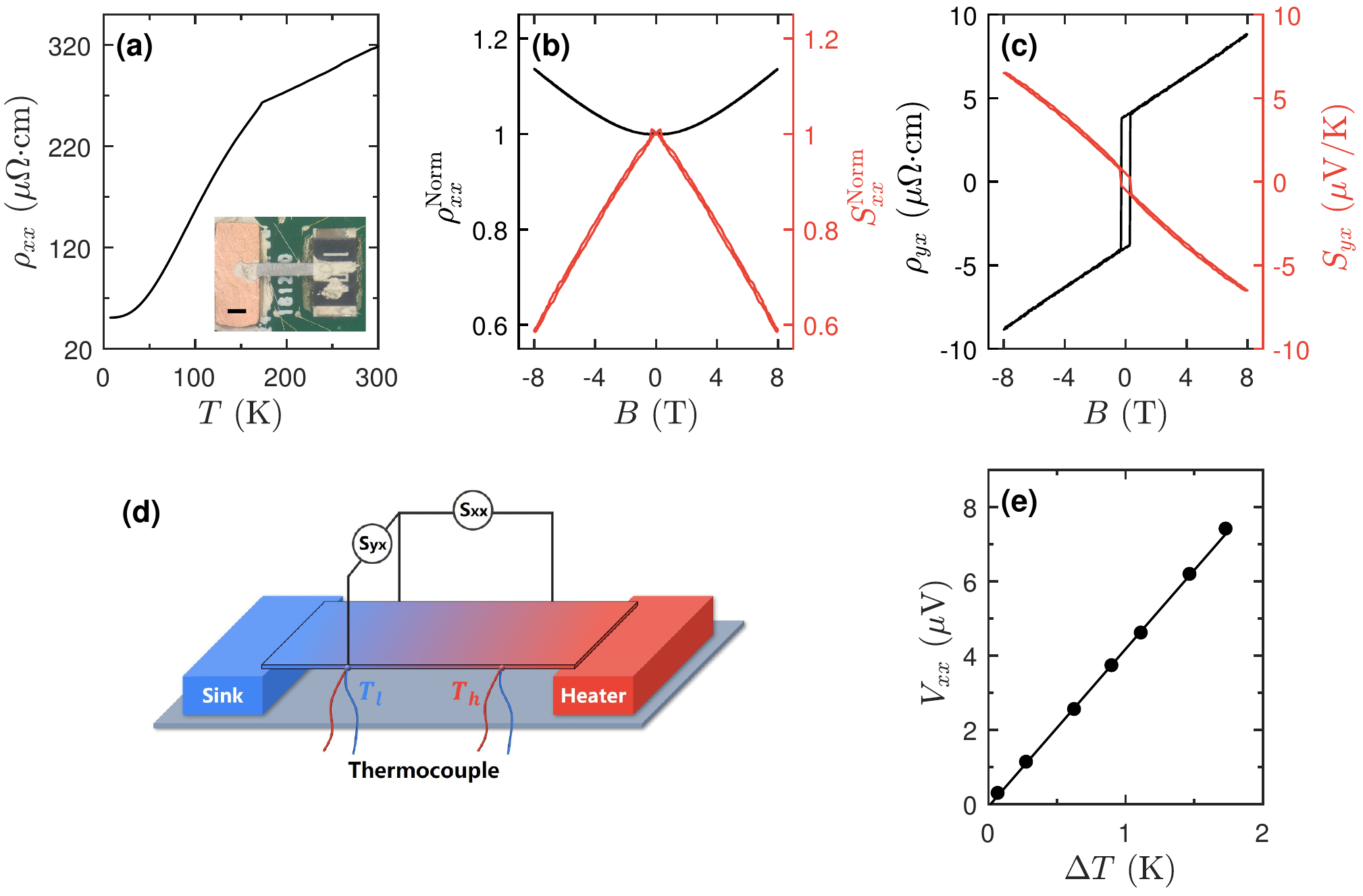}
			\caption{Basic transport properties of Co$_3$Sn$_2$S$_2$ crystal and thermoelectric measurement setup.  (a) Temperature dependence of longitudinal resistivity. A kink is observed at the Curie temperature $T_{c} = 175$ K. The inset is an optical image of the thermoelectric measurement setup. The scale bar is 1 mm. (b) Normalized longitudinal resistivity $\rho_{xx}$ (black) and Seebeck coefficient $S_{xx}$ (red) as a function of the perpendicular magnetic field at $T = 20$ K. The positive parabolic MR implies a dominant classic mechanism, while a negative and almost linear Seebeck may be attributed to the magnon-drag effect. (c) Anomalous Hall effect (black) and anomalous Nernst effect (red) at $T = 20$ K. (d) Diagram of thermoelectric measurement setup. Two thermocouples are attached at the edge of the sample. (e) Longitudinal thermoelectric signal $V_{xx}$ as a function of the temperature difference $\Delta T$. A linear dependence verifies the validation of our thermoelectric measurements.}
			\label{FigBasic}
		\end{center}
	\end{figure}

	Co$_3$Sn$_2$S$_2$ crystals were grown by a flux method. The typical size is about 3 mm $\times$ 3 mm $\times$ 1 mm. The crystals were polished to a bar shape for transport measurements. The sample presented in this work has a geometry of $\sim$ 3500 $\mu$m $\times$ 700 $\mu$m $\times$ 70 $\mu$m. To carry out thermoelectric measurements, one end of the sample was attached to a resistive heater using silver paste while the other end to a copper block, serving as a  heat sink. Two Type-E thermocouples were stuck to the sample to measure the temperature difference [see the inset of \rfig{FigBasic}(a) for the measurement setup and \rfig{FigBasic}(d) for a schematic diagram]. The temperature gradient was always along the $x$ axis (the [100] direction). The linear dependence of the longitudinal thermoelectric signal $V_{xx}$ on the temperature difference $\Delta T$ was verified, as shown in \rfig{FigBasic}(e). A typical current of 1.5 {\textendash} 2 mA was used, generating a temperature difference of 0.6 {\textendash} 1.1 K.
	
	The sample exhibits transport properties similar to previous results\cite{Jiang2021Jun,Wang2018,Guin2019a}, shown in \rfig{FigBasic}(a) {\textendash} (c). A clear kink in the temperature dependent resistivity [\rfig{FigBasic}(a)] indicates a ferromagnetic phase transition at $T_c \sim 175$ K\cite{Liu2018}. The normalized longitudinal resistivity $\rho_{xx}^\mathrm{Norm}$ and Seebeck coefficient $S_{xx}^\mathrm{Norm}$ at $T= 20$ K are shown in \rfig{FigBasic}(b) as a function of the magnetic field $B$. $\rho_{xx}$ displays a parabolic dependence on the magnetic field. On the other hand, $S_{xx}$ displays a negative and almost linear dependence on $B$. This behavior is attributed to the magnon-drag effect\cite{Costache2012,Raquet2002}, in which magnons flow against the temperature gradient and interacts with electrons, resulting in a thermoelectric voltage. Below the Curie temperature, both the anomalous Hall effect and the anomalous Nernst effect can be seen, manifested as a jump at the coercive field $B_c$, as shown in \rfig{FigBasic}(c). The Hall resistivity $\rho_{yx}$ above $B_c$ is more or less linear in the field. From the slope of the linear dependence, we obtain the ordinary Hall effect coefficient $R_\mathrm{H} = 0.61$ $\mu\Omega\cdot$cm/T and a corresponding carrier density of $1.0\times 10^{21}$ cm$^{-3}$. 
	
\begin{figure}[htbp]
		\begin{center}
			\includegraphics[width=1\columnwidth]{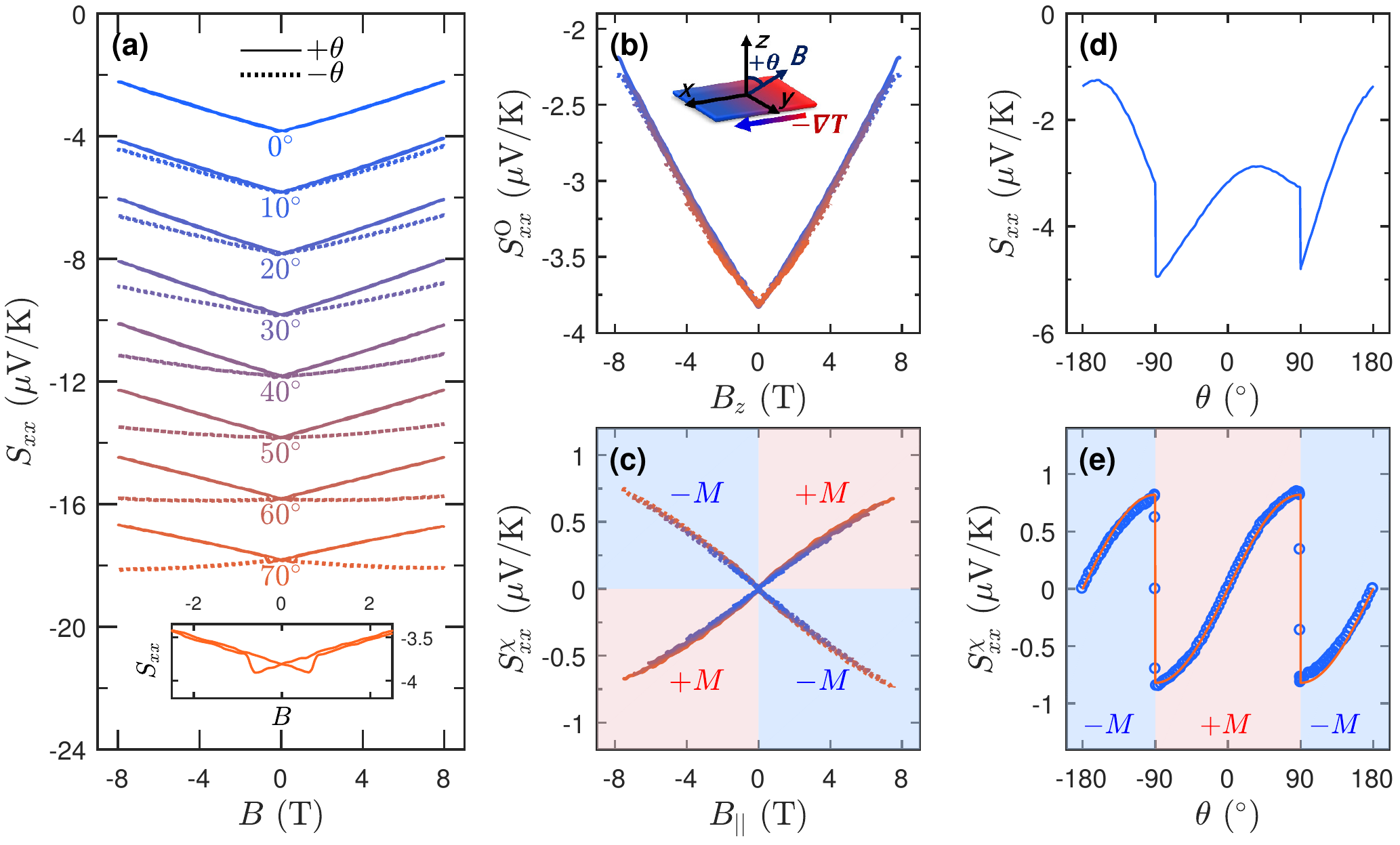}
			\caption{$B$- and $M$-antisymmetric Seebeck coefficient when $\vb{B}\perp \hat{x}$. (a) $S_{xx}$ as a function of the magnetic field at different angles. Data are shifted for clarity. Traces at $\pm\theta$ diverge with increasing field and angle. The inset shows the field dependence of $S_{xx}$ at $\theta=70^\circ$ at the low field region. Below $B_{c}$, the slope of the linear dependence shows opposite sign for $\pm M$. (b) $S_{xx}^\mathrm{O}$ as a function of $B_z$ at different angles. The inset depicts the measurement setup and the angle definition. (c) $S_{xx}^\chi$ as a function of $B_{\parallel}$. For positive (solid line) and negative (dashed line) magnetization, the slopes display opposite signs. (d) Angular dependence of $S_{xx}$ at $B=8$ T. There exists a pronounced asymmetry with respect to $\theta=0$. Because of the contact misalignment, there is a component of $S_{yx}$. Since $S_{yx}$ is nearly symmetric in $\theta$ when $B\perp x$, as seen in Fig.~S7, it does not affect the extracted $S^\chi_{xx}$. It is evident that $S_{xx}^\chi$ in (e) is very close to that in Fig.~S7(e)\cite{noteTEP}, which is extracted after removing the $S_{yx}$. (e) $S_{xx}^\chi$ versus $\theta$. $S_{xx}^\chi$ is obtained by $\theta$ antisymmetrizing the data in (d). Open circles are experimental data and the solid lines are best fits to $\sin\theta$.}
			\label{FigSbk}
		\end{center}
	\end{figure}

	Systematic measurements on the magnetic field dependence of the thermoelectric effect at different field directions were carried out. \rfig{FigSbk} shows the data at $T = 20$ K. When $B$ is along the $z$ axis, a negative and $B$-linear Seebeck coefficient is observed. When the magnetic field is rotated from the $z$ axis to the $y$ axis by an angle of $\theta$, depicted in the inset of \rfig{FigSbk}(b), $S_{xx}$ at $+\theta$ and $-\theta$ begin to diverge. The difference increases with $\theta$, seen in \rfig{FigSbk}(a). At $\pm\theta$, the perpendicular fields $B_{z}=B\cos\theta$ are the same, while the in-plane fields $B_{\parallel}=B\sin\theta$ have the same amplitude, but opposite signs. Therefore, the difference seemingly arises from $B_{\parallel}$ and depends on its polarity (antisymmetric). Let us tentatively postulate that $S_{xx}$ can be divided into two parts, $S_{xx}=S_{xx}^\mathrm{O}(B_{z})+S_{xx}^\chi(B_{\parallel})$, in which $S_{xx}^\mathrm{O}$ depends only on the perpendicular field, while $S_{xx}^\chi$ antisymmetrically depends on the in-plane field. Taking into account the possible dependence on magnetization $M$, $S_{xx}$ at $\pm\theta$ can be expressed as
	\begin{equation}
		\begin{cases}
			S_{xx}(+\theta,\pm B) = S_{xx}^\mathrm{O}(\pm M, \pm B_{z})+S_{xx}^\chi(\pm M, \pm B_{\parallel})  \\
			S_{xx}(-\theta,\pm B) = S_{xx}^\mathrm{O}(\pm M, \pm B_{z})+S_{xx}^\chi(\pm M, \mp B_{\parallel}).
		\end{cases}
		\label{eq.symmetry}
	\end{equation}
	Here we only consider regions in which $M$ aligns with $B$, i.e., sweeps from a high field to zero. Consequently, $S_{xx}^\mathrm{O}(M, B_{z})$ and $S_{xx}^\chi(M, B_{\parallel})$ can be obtained by symmetrizing and antisymmetrizing the data at $\pm\theta$, i.e., $S_{xx}^\mathrm{O}(\pm M, B_{z})=\left[S_{xx}(+\theta,\pm B)+S_{xx}(-\theta,\pm B)\right]/2$ and $S_{xx}^\chi(\pm M, B_{\parallel})=\left[S_{xx}(+\theta, \pm B)-S_{xx}(-\theta, \pm B)\right]/2$. In \rfig{FigSbk}(b), we plot $S_{xx}^\mathrm{O}$ as a function of $B_z$. All data at different angles collapse onto a single curve and are independent of $M$, which further confirms that $S_{xx}^\mathrm{O}$ is only determined by $B_{z}$ and verifies the validity of our decomposing method. $S_{xx}^\chi$ as a function of $B_{\parallel}$ at different angles are illustrated in \rfig{FigSbk}(c). All traces are scaled onto two curves, one for $+M$ and the other for $-M$. The two slopes are of opposite sign, indicating $S_{xx}^\chi(M, B_{\parallel})=-S_{xx}^\chi(-M, B_{\parallel})$. The excellent scaling of both $S_{xx}^\mathrm{O}$ and $S_{xx}^\chi$ strongly supports our postulation. Therefore, it can be concluded that $S_{xx}^\chi$ is antisymmetric in both $B$ and $M$, 
	\begin{equation}
		S_{xx}^\chi(M, B_{\parallel})=-S_{xx}^\chi(-M, B_{\parallel})=-S_{xx}^\chi(M, -B_{\parallel})=S_{xx}^\chi(-M, -B_{\parallel}).
		\label{eq.antisymmetry}
	\end{equation}
	In fact, this symmetry can also be directly seen at large $\theta$, where the out-of-plane field is strongly suppressed. In the inset of \rfig{FigSbk}(a), $S_{xx}$ at $\theta=70^\circ$ is zoomed in at the low field region. Below $B_c$, where $M$ remains unchanged, $S_{xx}$ is linear and antisymmetric in $B$ and the slope is reversed with $M$.
	
	Having revealed the dependence of $S_{xx}^\chi$ on $M$ and $B_{\parallel}$, we carry out the angular dependence measurement at $B = 8$ T. At this field, the magnetization is well saturated and mostly along the $z$ axis because of the strong magnetic crystalline anisotropy\cite{Liu2018,Shen2019}. The magnetic structure is of a single domain, unless $\theta$ is very close to $\pm 90^\circ$. When the magnetic field is rotated in the $yz$ plane ($\vb{B}\perp \hat{x}$), $S_{xx}$ displays a clear asymmetry with respect to $\theta=0$, as shown in \rfig{FigSbk}(d). The broken symmetry originates from the in-plane field induced $S_{xx}^\chi$. Similarly, the angular dependence of $S_{xx}^\chi$ is obtained by $\theta$ antisymmetrizing $S_{xx}(\theta)$ as $S_{xx}^\chi=\left[S_{xx}(+\theta,B)-S_{xx}(-\theta,B)\right]/2$. The resultant $S_{xx}^\chi$ can be well fit to a sine function in the region of $-90^\circ<\theta <+90^\circ$ where $M$ aligns with the $+z$ axis, which confirms that only the in-plane component of the field contributes to $S_{xx}^\chi$. $S_{xx}^\chi$ reverses its sign as $M$ flips at $\theta=\pm90^\circ$. In contrast, when $\vb{B}\perp \hat{y}$, $S_{xx}^\chi$ is much smaller (see Fig.~S6 in the Supplemental Material\cite{noteTEP}), suggesting that $S_{xx}^\chi$ only depends on $B_y$.
	
\begin{figure}[htbp]
		\begin{center}
			\includegraphics[width=1\columnwidth]{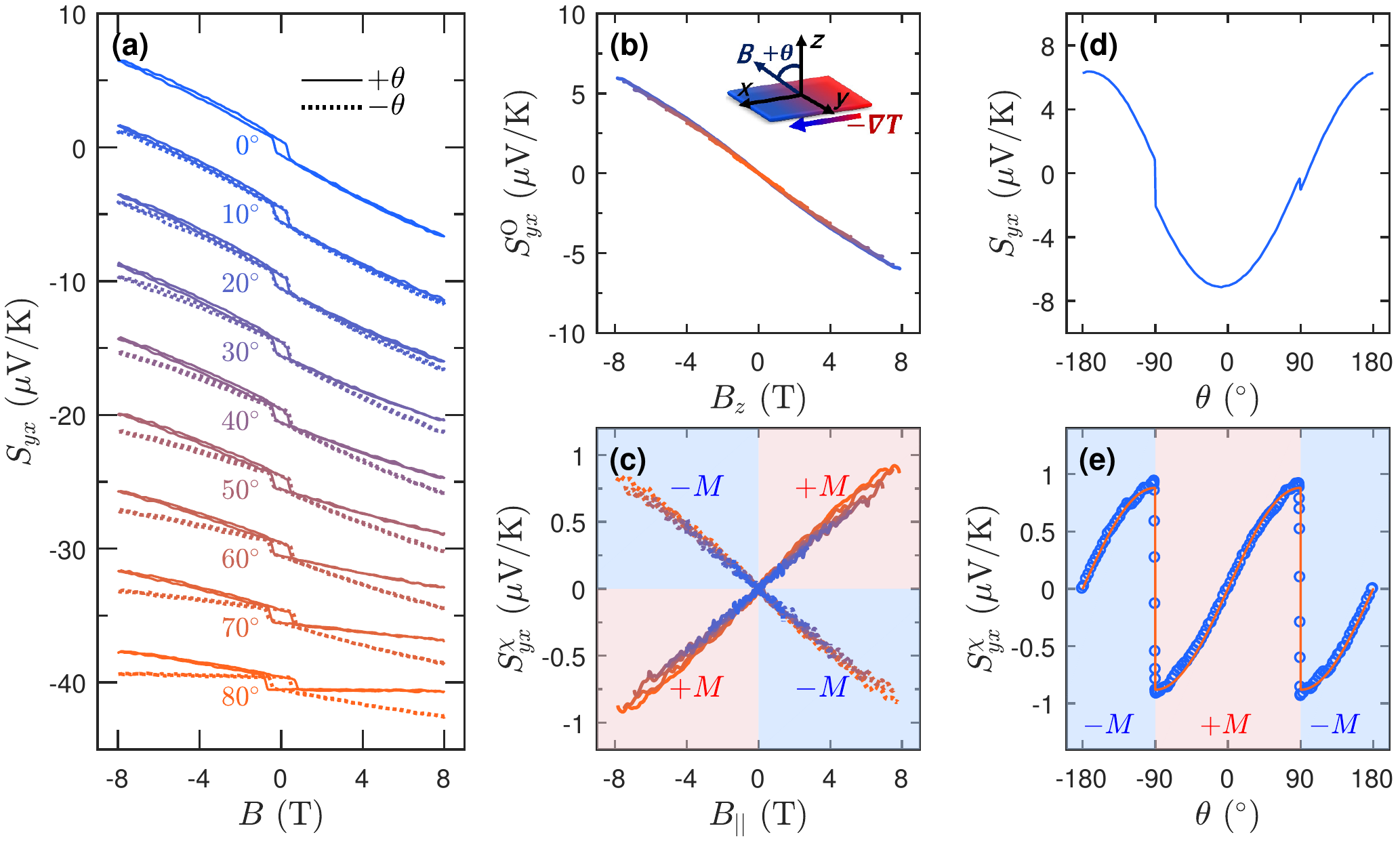}
			\caption{$B$- and $M$-antisymmetric Nernst effect when $\vb{B}\perp \hat{y}$. (a) Field dependence of Nernst signals at different angles. Data are shifted for clarity. Traces at $\pm\theta$ diverge with increasing field and angle. (b) $S_{yx}^\mathrm{O}$ versus $B_z$ at different angles. The inset depicts the rotation angle. (c) $S_{yx}^\chi$ versus $B_{\parallel}$. The slope of the field dependence changes sign with $M$. (d) Angular dependence of $S_{yx}$ at $B=8$ T. There exists a pronounced asymmetry with respect to $\theta=0$. Because of the contact misalignment, there is a slight component of $S_{xx}$. Moreover, it is nearly symmetric in $\theta$ when $B\perp y$, as seen in Fig.~S6\cite{noteTEP}. Therefore, it does not affect the extracted $S^\chi_{yx}$. (e) $S_{yx}^\chi$ versus $\theta$. $S_{yx}^\chi$ is obtained by $\theta$ antisymmetrizing the data in (d). Open circles are experimental data and the solid lines are best fits to $\sin\theta$. }
			\label{FigNst}
		\end{center}
	\end{figure}

	Although the unusual $S_{xx}^\chi$ emerges only when $\vb{B}\perp \hat{x}$, a nearly identical behavior appears in the Nernst signal $S_{yx}$ when $B$ is rotated in the $xz$ plane ($\vb{B}\perp \hat{y}$), as shown in the inset of \rfig{FigNst}(b). As $\theta$ increases, a marked difference between $+\theta$ and $-\theta$ develops. Following the preceding analysis and $\theta$-(anti)symmetrizing procedure, we decompose $S_{yx}$ into three parts, $S_{yx}^\mathrm{A}$, $S_{yx}^\mathrm{O}$ and $S_{yx}^\chi$. The additional term $S_{yx}^\mathrm{A}$ denotes the anomalous Nernst effect, which is assumed to be a constant, as the magnetization of Co$_3$Sn$_2$S$_2$ stays almost unchanged above $B_{c}$ at 20 K\cite{Guin2019a}. The extracted $S_{yx}^\mathrm{O}$ at different field angles are plotted as a function of $B_z$ in \rfig{FigNst}(b). All traces overlap, indicating that $S_{yx}^\mathrm{O}$ depends solely on $B_z$. $S_{yx}^\chi$ at different $\theta$ collapse onto a single curve, when they are plotted against $B_{\parallel}$. Just like $S_{xx}^\chi$ when $\vb{B}\perp \hat{x}$, here $S_{yx}^\chi$ is linear in $B_{\parallel}$ and reverses its sign as $M_{z}$ switches polarity. Again, the perfect scaling of $S_{yx}^\mathrm{O}$ and $S_{yx}^\chi$ validates our method of decomposing $S_{yx}$. The angular dependence of $S_{yx}$ also exhibits a substantial asymmetry [\rfig{FigNst}(d)], from which a significant $S_{yx}^\chi$ is extracted. It follows a sine dependence, indicating that only the in-plane component of the field contributes. In addition, no significant $S_{yx}^\chi$ was observed when $\vb{B}\perp \hat{x}$ (see Fig.~S7 in the Supplemental Material\cite{noteTEP}), so it can be concluded that $S_{yx}^{\chi}$ only depends on $B_{x}$.
	
	Our main experimental observations can be summarized as an additional Seebeck signal when $\vb{B}\perp \hat{x}\ (\nabla T)$ and an additional Nernst signal when $\vb{B}\perp \hat{y}$. Both signals are antisymmetric in $B$ and $M$. Similar electrical transport effects have been investigated in a quasiclassical Boltzmann framework\cite{Zyuzin2017,Sharma2017,Das2019,Ma2019a}. Based on equations of motion with Berry curvature effects, three $B$-linear currents driven by an electric field $E$ have been identified. They originate from the Berry curvature induced anomalous velocity $\frac{e}{\hbar}(\vb{v_k\cdot \Omega_k})\vb{B}$, the chemical potential imbalance $(\vb{B\cdot E})(\vb{v_k\cdot \Omega_k})$ between Weyl nodes of opposite chiralities, and the Berry curvature correction to the phase space volume $[1+\frac{e}{\hbar}(\vb{B\cdot\Omega_k})]^{-1}$\cite{Das2019,Ma2019a}. Here, $\vb{v_k}$ and $\vb{\Omega_k}$ denote the group velocity and the Berry curvature, respectively. Without a tilt, the integrals of these terms over the Fermi surface vanish. They are nonzero only when Weyl cones of opposite chiralities tilt to opposite directions. The first two terms are associated with a so-called one-dimensional chiral anomaly\cite{Zyuzin2017}. These currents were recently confirmed by experiments on Co$_3$Sn$_2$S$_2$\cite{Jiang2021Jun}. 
	
	The $B$-linear terms in electrical transport result from the Berry curvature, which goes into the equations of motion via an anomalous velocity, $\frac{e}{\hbar}\vb{E}\times\vb{\Omega}$. However, this velocity is absent in thermoelectric effects as the driving force is a statistic force, $\nabla T$, instead of a mechanical force $E$ in the case of electrical transport. It is not clear whether these currents will appear in thermoelectric effects. It has been shown that the correction to orbital magnetization due to the Berry curvature needs to be considered in thermoelectric effects\cite{Xiao2006}. In the presence of a temperature gradient, the orbital magnetization gives rise to an extra term in the current, which plays a similar role as the anomalous velocity under electric field. Zyuzin has calculated the thermoelectric effect in tilted Weyl semimetals, taken into account the orbital magnetic moment, and found $B$-linear contributions similar to those in electrical conductivity\cite{Zyuzin2017}. A calculation using the Mott relation yields consistent results\cite{Das2019a}.
	
	Our observation can be qualitatively understood based on these theories. Note that the Weyl nodes in Co$_3$Sn$_2$S$_2$ are tilted in the $y$ direction. When the in-plane component of $B$ is parallel to the $x$ axis, it generates a difference in the chemical potential between Weyl nodes of opposite chiralities, $\propto \vb{B}\cdot \nabla T$. This chiral chemical potential, combined with a tilted Dirac dispersion, gives rise to a current along the $y$ axis, contributing to the Nernst effect. When the in-plane component of $B$ is parallel to the $y$ axis, the phase space correction term will give rise to a current along the $x$ axis, contributing to the Seebeck effect. When the magnetization is reversed, Weyl nodes flip their chirality. Hence, the current is reversed.
	
	We perform a theoretical calculation based on the semiclassical Boltzmann equation. We employ a two-band Hamiltonian of the tilted Weyl semimetal and take into account the Berry curvature correction and orbital magnetization. The detailed theoretical calculations can be found in the Supplemental Material\cite{noteTEP}. The parameters for the band are taken from experimental studies\cite{Liu2018}. The key parameter, tilt of bands, is obtained by fitting our model with the experimental results of the chirality dependent Hall effect and the antisymmetric magnetoresistance\cite{Jiang2021Jun}. After considering the $C_3$ symmetry of Co$_3$Sn$_2$S$_2$, the calculated $B$-antisymmetric Seebeck and Nernst coefficient are qualitatively consistent with our experimental results in sign and magnitude. As the magnetization reverses, the calculated thermoelectric coefficient changes its sign, which also agrees with our experiments.
	
	In conclusion, we have uncovered exotic Seebeck and Nernst effects in a tilted Weyl semimetal. In particular, the Seebeck effect is antisymmetric in the magnetic field. Both the Seebeck effect and the Nernst effect are also antisymmetric in magnetization. It is shown that these thermoelectric phenomena arise from the tilting of Weyl cones.
	
	\begin{acknowledgements}
		This work was mainly supported by National Key R\&D Program of China (No. 2020YFA0308800) and NSFC (No. 12074009, No. 11774009). E.L. and S.Z. acknowledge the financial support from National Key R\&D Program of China (No. 2019YFA0704900), Fundamental Science Centre of the National Natural Science Foundation of China (No. 52088101) and the Key Research Program of the Chinese Academy of Sciences (No. ZDRW-CN-2021-3).
	\end{acknowledgements}
	

\begin{thebibliography}{54}%
		\makeatletter
		\providecommand \@ifxundefined [1]{%
			\@ifx{#1\undefined}
		}%
		\providecommand \@ifnum [1]{%
			\ifnum #1\expandafter \@firstoftwo
			\else \expandafter \@secondoftwo
			\fi
		}%
		\providecommand \@ifx [1]{%
			\ifx #1\expandafter \@firstoftwo
			\else \expandafter \@secondoftwo
			\fi
		}%
		\providecommand \natexlab [1]{#1}%
		\providecommand \enquote  [1]{``#1''}%
		\providecommand \bibnamefont  [1]{#1}%
		\providecommand \bibfnamefont [1]{#1}%
		\providecommand \citenamefont [1]{#1}%
		\providecommand \href@noop [0]{\@secondoftwo}%
		\providecommand \href [0]{\begingroup \@sanitize@url \@href}%
		\providecommand \@href[1]{\@@startlink{#1}\@@href}%
		\providecommand \@@href[1]{\endgroup#1\@@endlink}%
		\providecommand \@sanitize@url [0]{\catcode `\\12\catcode `\$12\catcode
			`\&12\catcode `\#12\catcode `\^12\catcode `\_12\catcode `\%12\relax}%
		\providecommand \@@startlink[1]{}%
		\providecommand \@@endlink[0]{}%
		\providecommand \url  [0]{\begingroup\@sanitize@url \@url }%
		\providecommand \@url [1]{\endgroup\@href {#1}{\urlprefix }}%
		\providecommand \urlprefix  [0]{URL }%
		\providecommand \Eprint [0]{\href }%
		\providecommand \doibase [0]{http://dx.doi.org/}%
		\providecommand \selectlanguage [0]{\@gobble}%
		\providecommand \bibinfo  [0]{\@secondoftwo}%
		\providecommand \bibfield  [0]{\@secondoftwo}%
		\providecommand \translation [1]{[#1]}%
		\providecommand \BibitemOpen [0]{}%
		\providecommand \bibitemStop [0]{}%
		\providecommand \bibitemNoStop [0]{.\EOS\space}%
		\providecommand \EOS [0]{\spacefactor3000\relax}%
		\providecommand \BibitemShut  [1]{\csname bibitem#1\endcsname}%
		\let\auto@bib@innerbib\@empty
		\bibitem [{\citenamefont {Wan}\ \emph {et~al.}(2011)\citenamefont {Wan},
			\citenamefont {Turner}, \citenamefont {Vishwanath},\ and\ \citenamefont
			{Savrasov}}]{Wan2011}%
		\BibitemOpen
		\bibfield  {author} {\bibinfo {author} {\bibfnamefont {X.}~\bibnamefont
				{Wan}}, \bibinfo {author} {\bibfnamefont {A.~M.}\ \bibnamefont {Turner}},
			\bibinfo {author} {\bibfnamefont {A.}~\bibnamefont {Vishwanath}}, \ and\
			\bibinfo {author} {\bibfnamefont {S.~Y.}\ \bibnamefont {Savrasov}},\ }\href
		{\doibase 10.1103/PhysRevB.83.205101} {\bibfield  {journal} {\bibinfo
				{journal} {Phys. Rev. B}\ }\textbf {\bibinfo {volume} {83}},\ \bibinfo
			{pages} {205101} (\bibinfo {year} {2011})}\BibitemShut {NoStop}%
		\bibitem [{\citenamefont {Weng}\ \emph {et~al.}(2015)\citenamefont {Weng},
			\citenamefont {Fang}, \citenamefont {Fang}, \citenamefont {Bernevig},\ and\
			\citenamefont {Dai}}]{Weng2015Mar}%
		\BibitemOpen
		\bibfield  {author} {\bibinfo {author} {\bibfnamefont {H.}~\bibnamefont
				{Weng}}, \bibinfo {author} {\bibfnamefont {C.}~\bibnamefont {Fang}}, \bibinfo
			{author} {\bibfnamefont {Z.}~\bibnamefont {Fang}}, \bibinfo {author}
			{\bibfnamefont {B.~A.}\ \bibnamefont {Bernevig}}, \ and\ \bibinfo {author}
			{\bibfnamefont {X.}~\bibnamefont {Dai}},\ }\href {\doibase
			10.1103/PhysRevX.5.011029} {\bibfield  {journal} {\bibinfo  {journal} {Phys.
					Rev. X}\ }\textbf {\bibinfo {volume} {5}},\ \bibinfo {pages} {011029}
			(\bibinfo {year} {2015})}\BibitemShut {NoStop}%
		\bibitem [{\citenamefont {Huang}\ \emph {et~al.}(2015)\citenamefont {Huang},
			\citenamefont {Xu}, \citenamefont {Belopolski}, \citenamefont {Lee},
			\citenamefont {Chang}, \citenamefont {Wang}, \citenamefont {Alidoust},
			\citenamefont {Bian}, \citenamefont {Neupane}, \citenamefont {Zhang},
			\citenamefont {Jia}, \citenamefont {Bansil}, \citenamefont {Lin},\ and\
			\citenamefont {Hasan}}]{Huang2015Jun}%
		\BibitemOpen
		\bibfield  {author} {\bibinfo {author} {\bibfnamefont {S.-M.}\ \bibnamefont
				{Huang}}, \bibinfo {author} {\bibfnamefont {S.-Y.}\ \bibnamefont {Xu}},
			\bibinfo {author} {\bibfnamefont {I.}~\bibnamefont {Belopolski}}, \bibinfo
			{author} {\bibfnamefont {C.-C.}\ \bibnamefont {Lee}}, \bibinfo {author}
			{\bibfnamefont {G.}~\bibnamefont {Chang}}, \bibinfo {author} {\bibfnamefont
				{B.}~\bibnamefont {Wang}}, \bibinfo {author} {\bibfnamefont {N.}~\bibnamefont
				{Alidoust}}, \bibinfo {author} {\bibfnamefont {G.}~\bibnamefont {Bian}},
			\bibinfo {author} {\bibfnamefont {M.}~\bibnamefont {Neupane}}, \bibinfo
			{author} {\bibfnamefont {C.}~\bibnamefont {Zhang}}, \bibinfo {author}
			{\bibfnamefont {S.}~\bibnamefont {Jia}}, \bibinfo {author} {\bibfnamefont
				{A.}~\bibnamefont {Bansil}}, \bibinfo {author} {\bibfnamefont
				{H.}~\bibnamefont {Lin}}, \ and\ \bibinfo {author} {\bibfnamefont {M.~Z.}\
				\bibnamefont {Hasan}},\ }\href {\doibase 10.1038/ncomms8373} {\bibfield
			{journal} {\bibinfo  {journal} {Nat. Commun.}\ }\textbf {\bibinfo {volume}
				{6}},\ \bibinfo {pages} {1} (\bibinfo {year} {2015})}\BibitemShut {NoStop}%
		\bibitem [{\citenamefont {Lv}\ \emph {et~al.}(2015{\natexlab{a}})\citenamefont
			{Lv}, \citenamefont {Weng}, \citenamefont {Fu}, \citenamefont {Wang},
			\citenamefont {Miao}, \citenamefont {Ma}, \citenamefont {Richard},
			\citenamefont {Huang}, \citenamefont {Zhao}, \citenamefont {Chen},
			\citenamefont {Fang}, \citenamefont {Dai}, \citenamefont {Qian},\ and\
			\citenamefont {Ding}}]{Lv2015Jul}%
		\BibitemOpen
		\bibfield  {author} {\bibinfo {author} {\bibfnamefont {B.~Q.}\ \bibnamefont
				{Lv}}, \bibinfo {author} {\bibfnamefont {H.~M.}\ \bibnamefont {Weng}},
			\bibinfo {author} {\bibfnamefont {B.~B.}\ \bibnamefont {Fu}}, \bibinfo
			{author} {\bibfnamefont {X.~P.}\ \bibnamefont {Wang}}, \bibinfo {author}
			{\bibfnamefont {H.}~\bibnamefont {Miao}}, \bibinfo {author} {\bibfnamefont
				{J.}~\bibnamefont {Ma}}, \bibinfo {author} {\bibfnamefont {P.}~\bibnamefont
				{Richard}}, \bibinfo {author} {\bibfnamefont {X.~C.}\ \bibnamefont {Huang}},
			\bibinfo {author} {\bibfnamefont {L.~X.}\ \bibnamefont {Zhao}}, \bibinfo
			{author} {\bibfnamefont {G.~F.}\ \bibnamefont {Chen}}, \bibinfo {author}
			{\bibfnamefont {Z.}~\bibnamefont {Fang}}, \bibinfo {author} {\bibfnamefont
				{X.}~\bibnamefont {Dai}}, \bibinfo {author} {\bibfnamefont {T.}~\bibnamefont
				{Qian}}, \ and\ \bibinfo {author} {\bibfnamefont {H.}~\bibnamefont {Ding}},\
		}\href {\doibase 10.1103/PhysRevX.5.031013} {\bibfield  {journal} {\bibinfo
				{journal} {Phys. Rev. X}\ }\textbf {\bibinfo {volume} {5}},\ \bibinfo {pages}
			{031013} (\bibinfo {year} {2015}{\natexlab{a}})}\BibitemShut {NoStop}%
		\bibitem [{\citenamefont {Xu}\ \emph {et~al.}(2015{\natexlab{a}})\citenamefont
			{Xu}, \citenamefont {Belopolski}, \citenamefont {Alidoust}, \citenamefont
			{Neupane}, \citenamefont {Bian}, \citenamefont {Zhang}, \citenamefont
			{Sankar}, \citenamefont {Chang}, \citenamefont {Yuan}, \citenamefont {Lee},
			\citenamefont {Huang}, \citenamefont {Zheng}, \citenamefont {Ma},
			\citenamefont {Sanchez}, \citenamefont {Wang}, \citenamefont {Bansil},
			\citenamefont {Chou}, \citenamefont {Shibayev}, \citenamefont {Lin},
			\citenamefont {Jia},\ and\ \citenamefont {Hasan}}]{Xu2015Aug}%
		\BibitemOpen
		\bibfield  {author} {\bibinfo {author} {\bibfnamefont {S.-Y.}\ \bibnamefont
				{Xu}}, \bibinfo {author} {\bibfnamefont {I.}~\bibnamefont {Belopolski}},
			\bibinfo {author} {\bibfnamefont {N.}~\bibnamefont {Alidoust}}, \bibinfo
			{author} {\bibfnamefont {M.}~\bibnamefont {Neupane}}, \bibinfo {author}
			{\bibfnamefont {G.}~\bibnamefont {Bian}}, \bibinfo {author} {\bibfnamefont
				{C.}~\bibnamefont {Zhang}}, \bibinfo {author} {\bibfnamefont
				{R.}~\bibnamefont {Sankar}}, \bibinfo {author} {\bibfnamefont
				{G.}~\bibnamefont {Chang}}, \bibinfo {author} {\bibfnamefont
				{Z.}~\bibnamefont {Yuan}}, \bibinfo {author} {\bibfnamefont {C.-C.}\
				\bibnamefont {Lee}}, \bibinfo {author} {\bibfnamefont {S.-M.}\ \bibnamefont
				{Huang}}, \bibinfo {author} {\bibfnamefont {H.}~\bibnamefont {Zheng}},
			\bibinfo {author} {\bibfnamefont {J.}~\bibnamefont {Ma}}, \bibinfo {author}
			{\bibfnamefont {D.~S.}\ \bibnamefont {Sanchez}}, \bibinfo {author}
			{\bibfnamefont {B.}~\bibnamefont {Wang}}, \bibinfo {author} {\bibfnamefont
				{A.}~\bibnamefont {Bansil}}, \bibinfo {author} {\bibfnamefont
				{F.}~\bibnamefont {Chou}}, \bibinfo {author} {\bibfnamefont {P.~P.}\
				\bibnamefont {Shibayev}}, \bibinfo {author} {\bibfnamefont {H.}~\bibnamefont
				{Lin}}, \bibinfo {author} {\bibfnamefont {S.}~\bibnamefont {Jia}}, \ and\
			\bibinfo {author} {\bibfnamefont {M.~Z.}\ \bibnamefont {Hasan}},\ }\href
		{https://www.science.org/doi/10.1126/science.aaa9297} {\bibfield  {journal}
			{\bibinfo  {journal} {Science}\ } (\bibinfo {year}
			{2015}{\natexlab{a}})}\BibitemShut {NoStop}%
		\bibitem [{\citenamefont {Lv}\ \emph {et~al.}(2015{\natexlab{b}})\citenamefont
			{Lv}, \citenamefont {Xu}, \citenamefont {Weng}, \citenamefont {Ma},
			\citenamefont {Richard}, \citenamefont {Huang}, \citenamefont {Zhao},
			\citenamefont {Chen}, \citenamefont {Matt}, \citenamefont {Bisti},
			\citenamefont {Strocov}, \citenamefont {Mesot}, \citenamefont {Fang},
			\citenamefont {Dai}, \citenamefont {Qian}, \citenamefont {Shi},\ and\
			\citenamefont {Ding}}]{Lv2015Sep}%
		\BibitemOpen
		\bibfield  {author} {\bibinfo {author} {\bibfnamefont {B.~Q.}\ \bibnamefont
				{Lv}}, \bibinfo {author} {\bibfnamefont {N.}~\bibnamefont {Xu}}, \bibinfo
			{author} {\bibfnamefont {H.~M.}\ \bibnamefont {Weng}}, \bibinfo {author}
			{\bibfnamefont {J.~Z.}\ \bibnamefont {Ma}}, \bibinfo {author} {\bibfnamefont
				{P.}~\bibnamefont {Richard}}, \bibinfo {author} {\bibfnamefont {X.~C.}\
				\bibnamefont {Huang}}, \bibinfo {author} {\bibfnamefont {L.~X.}\ \bibnamefont
				{Zhao}}, \bibinfo {author} {\bibfnamefont {G.~F.}\ \bibnamefont {Chen}},
			\bibinfo {author} {\bibfnamefont {C.~E.}\ \bibnamefont {Matt}}, \bibinfo
			{author} {\bibfnamefont {F.}~\bibnamefont {Bisti}}, \bibinfo {author}
			{\bibfnamefont {V.~N.}\ \bibnamefont {Strocov}}, \bibinfo {author}
			{\bibfnamefont {J.}~\bibnamefont {Mesot}}, \bibinfo {author} {\bibfnamefont
				{Z.}~\bibnamefont {Fang}}, \bibinfo {author} {\bibfnamefont {X.}~\bibnamefont
				{Dai}}, \bibinfo {author} {\bibfnamefont {T.}~\bibnamefont {Qian}}, \bibinfo
			{author} {\bibfnamefont {M.}~\bibnamefont {Shi}}, \ and\ \bibinfo {author}
			{\bibfnamefont {H.}~\bibnamefont {Ding}},\ }\href {\doibase
			10.1038/nphys3426} {\bibfield  {journal} {\bibinfo  {journal} {Nat. Phys.}\
			}\textbf {\bibinfo {volume} {11}},\ \bibinfo {pages} {724} (\bibinfo {year}
			{2015}{\natexlab{b}})}\BibitemShut {NoStop}%
		\bibitem [{\citenamefont {Xu}\ \emph {et~al.}(2015{\natexlab{b}})\citenamefont
			{Xu}, \citenamefont {Alidoust}, \citenamefont {Belopolski}, \citenamefont
			{Yuan}, \citenamefont {Bian}, \citenamefont {Chang}, \citenamefont {Zheng},
			\citenamefont {Strocov}, \citenamefont {Sanchez}, \citenamefont {Chang},
			\citenamefont {Zhang}, \citenamefont {Mou}, \citenamefont {Wu}, \citenamefont
			{Huang}, \citenamefont {Lee}, \citenamefont {Huang}, \citenamefont {Wang},
			\citenamefont {Bansil}, \citenamefont {Jeng}, \citenamefont {Neupert},
			\citenamefont {Kaminski}, \citenamefont {Lin}, \citenamefont {Jia},\ and\
			\citenamefont {Zahid~Hasan}}]{Xu2015Sep}%
		\BibitemOpen
		\bibfield  {author} {\bibinfo {author} {\bibfnamefont {S.-Y.}\ \bibnamefont
				{Xu}}, \bibinfo {author} {\bibfnamefont {N.}~\bibnamefont {Alidoust}},
			\bibinfo {author} {\bibfnamefont {I.}~\bibnamefont {Belopolski}}, \bibinfo
			{author} {\bibfnamefont {Z.}~\bibnamefont {Yuan}}, \bibinfo {author}
			{\bibfnamefont {G.}~\bibnamefont {Bian}}, \bibinfo {author} {\bibfnamefont
				{T.-R.}\ \bibnamefont {Chang}}, \bibinfo {author} {\bibfnamefont
				{H.}~\bibnamefont {Zheng}}, \bibinfo {author} {\bibfnamefont {V.~N.}\
				\bibnamefont {Strocov}}, \bibinfo {author} {\bibfnamefont {D.~S.}\
				\bibnamefont {Sanchez}}, \bibinfo {author} {\bibfnamefont {G.}~\bibnamefont
				{Chang}}, \bibinfo {author} {\bibfnamefont {C.}~\bibnamefont {Zhang}},
			\bibinfo {author} {\bibfnamefont {D.}~\bibnamefont {Mou}}, \bibinfo {author}
			{\bibfnamefont {Y.}~\bibnamefont {Wu}}, \bibinfo {author} {\bibfnamefont
				{L.}~\bibnamefont {Huang}}, \bibinfo {author} {\bibfnamefont {C.-C.}\
				\bibnamefont {Lee}}, \bibinfo {author} {\bibfnamefont {S.-M.}\ \bibnamefont
				{Huang}}, \bibinfo {author} {\bibfnamefont {B.}~\bibnamefont {Wang}},
			\bibinfo {author} {\bibfnamefont {A.}~\bibnamefont {Bansil}}, \bibinfo
			{author} {\bibfnamefont {H.-T.}\ \bibnamefont {Jeng}}, \bibinfo {author}
			{\bibfnamefont {T.}~\bibnamefont {Neupert}}, \bibinfo {author} {\bibfnamefont
				{A.}~\bibnamefont {Kaminski}}, \bibinfo {author} {\bibfnamefont
				{H.}~\bibnamefont {Lin}}, \bibinfo {author} {\bibfnamefont {S.}~\bibnamefont
				{Jia}}, \ and\ \bibinfo {author} {\bibfnamefont {M.}~\bibnamefont
				{Zahid~Hasan}},\ }\href {\doibase 10.1038/nphys3437} {\bibfield  {journal}
			{\bibinfo  {journal} {Nat. Phys.}\ }\textbf {\bibinfo {volume} {11}},\
			\bibinfo {pages} {748} (\bibinfo {year} {2015}{\natexlab{b}})}\BibitemShut
		{NoStop}%
		\bibitem [{\citenamefont {Armitage}\ \emph {et~al.}(2018)\citenamefont
			{Armitage}, \citenamefont {Mele},\ and\ \citenamefont
			{Vishwanath}}]{Armitage2018}%
		\BibitemOpen
		\bibfield  {author} {\bibinfo {author} {\bibfnamefont {N.~P.}\ \bibnamefont
				{Armitage}}, \bibinfo {author} {\bibfnamefont {E.~J.}\ \bibnamefont {Mele}},
			\ and\ \bibinfo {author} {\bibfnamefont {A.}~\bibnamefont {Vishwanath}},\
		}\href {\doibase 10.1103/RevModPhys.90.015001} {\bibfield  {journal}
			{\bibinfo  {journal} {Rev. Mod. Phys.}\ }\textbf {\bibinfo {volume} {90}},\
			\bibinfo {pages} {015001} (\bibinfo {year} {2018})}\BibitemShut {NoStop}%
		\bibitem [{\citenamefont {Nagaosa}\ \emph {et~al.}(2020)\citenamefont
			{Nagaosa}, \citenamefont {Morimoto},\ and\ \citenamefont
			{Tokura}}]{Nagaosa2020}%
		\BibitemOpen
		\bibfield  {author} {\bibinfo {author} {\bibfnamefont {N.}~\bibnamefont
				{Nagaosa}}, \bibinfo {author} {\bibfnamefont {T.}~\bibnamefont {Morimoto}}, \
			and\ \bibinfo {author} {\bibfnamefont {Y.}~\bibnamefont {Tokura}},\ }\href
		{\doibase 10.1038/s41578-020-0208-y} {\bibfield  {journal} {\bibinfo
				{journal} {Nat. Rev. Mater.}\ }\textbf {\bibinfo {volume} {5}},\ \bibinfo
			{pages} {621} (\bibinfo {year} {2020})}\BibitemShut {NoStop}%
		\bibitem [{\citenamefont {Volovik}(2016)}]{Volovik2016Nov}%
		\BibitemOpen
		\bibfield  {author} {\bibinfo {author} {\bibfnamefont {G.~E.}\ \bibnamefont
				{Volovik}},\ }\href {\doibase 10.1134/S0021364016210050} {\bibfield
			{journal} {\bibinfo  {journal} {JETP Lett.}\ }\textbf {\bibinfo {volume}
				{104}},\ \bibinfo {pages} {645} (\bibinfo {year} {2016})}\BibitemShut
		{NoStop}%
		\bibitem [{\citenamefont {Carbotte}(2016)}]{Carbotte2016Oct}%
		\BibitemOpen
		\bibfield  {author} {\bibinfo {author} {\bibfnamefont {J.~P.}\ \bibnamefont
				{Carbotte}},\ }\href {\doibase 10.1103/PhysRevB.94.165111} {\bibfield
			{journal} {\bibinfo  {journal} {Phys. Rev. B}\ }\textbf {\bibinfo {volume}
				{94}},\ \bibinfo {pages} {165111} (\bibinfo {year} {2016})}\BibitemShut
		{NoStop}%
		\bibitem [{\citenamefont {Isobe}\ and\ \citenamefont
			{Nagaosa}(2016)}]{Isobe2016Mar}%
		\BibitemOpen
		\bibfield  {author} {\bibinfo {author} {\bibfnamefont {H.}~\bibnamefont
				{Isobe}}\ and\ \bibinfo {author} {\bibfnamefont {N.}~\bibnamefont
				{Nagaosa}},\ }\href {\doibase 10.1103/PhysRevLett.116.116803} {\bibfield
			{journal} {\bibinfo  {journal} {Phys. Rev. Lett.}\ }\textbf {\bibinfo
				{volume} {116}},\ \bibinfo {pages} {116803} (\bibinfo {year}
			{2016})}\BibitemShut {NoStop}%
		\bibitem [{\citenamefont {Weststr\"{o}m}\ and\ \citenamefont
			{Ojanen}(2017)}]{Weststrom2017Oct}%
		\BibitemOpen
		\bibfield  {author} {\bibinfo {author} {\bibfnamefont {A.}~\bibnamefont
				{Weststr\"{o}m}}\ and\ \bibinfo {author} {\bibfnamefont {T.}~\bibnamefont
				{Ojanen}},\ }\href {\doibase 10.1103/PhysRevX.7.041026} {\bibfield  {journal}
			{\bibinfo  {journal} {Phys. Rev. X}\ }\textbf {\bibinfo {volume} {7}},\
			\bibinfo {pages} {041026} (\bibinfo {year} {2017})}\BibitemShut {NoStop}%
		\bibitem [{\citenamefont {Farajollahpour}\ \emph {et~al.}(2019)\citenamefont
			{Farajollahpour}, \citenamefont {Faraei},\ and\ \citenamefont
			{Jafari}}]{Farajollahpour2019Jun}%
		\BibitemOpen
		\bibfield  {author} {\bibinfo {author} {\bibfnamefont {T.}~\bibnamefont
				{Farajollahpour}}, \bibinfo {author} {\bibfnamefont {Z.}~\bibnamefont
				{Faraei}}, \ and\ \bibinfo {author} {\bibfnamefont {S.~A.}\ \bibnamefont
				{Jafari}},\ }\href {\doibase 10.1103/PhysRevB.99.235150} {\bibfield
			{journal} {\bibinfo  {journal} {Phys. Rev. B}\ }\textbf {\bibinfo {volume}
				{99}},\ \bibinfo {pages} {235150} (\bibinfo {year} {2019})}\BibitemShut
		{NoStop}%
		\bibitem [{\citenamefont {Ferreiros}\ \emph {et~al.}(2019)\citenamefont
			{Ferreiros}, \citenamefont {Kedem}, \citenamefont {Bergholtz},\ and\
			\citenamefont {Bardarson}}]{Ferreiros2019Feb}%
		\BibitemOpen
		\bibfield  {author} {\bibinfo {author} {\bibfnamefont {Y.}~\bibnamefont
				{Ferreiros}}, \bibinfo {author} {\bibfnamefont {Y.}~\bibnamefont {Kedem}},
			\bibinfo {author} {\bibfnamefont {E.~J.}\ \bibnamefont {Bergholtz}}, \ and\
			\bibinfo {author} {\bibfnamefont {J.~H.}\ \bibnamefont {Bardarson}},\ }\href
		{\doibase 10.1103/PhysRevLett.122.056601} {\bibfield  {journal} {\bibinfo
				{journal} {Phys. Rev. Lett.}\ }\textbf {\bibinfo {volume} {122}},\ \bibinfo
			{pages} {056601} (\bibinfo {year} {2019})}\BibitemShut {NoStop}%
		\bibitem [{\citenamefont {Ferreiros}\ \emph {et~al.}(2017)\citenamefont
			{Ferreiros}, \citenamefont {Zyuzin},\ and\ \citenamefont
			{Bardarson}}]{Ferreiros2017}%
		\BibitemOpen
		\bibfield  {author} {\bibinfo {author} {\bibfnamefont {Y.}~\bibnamefont
				{Ferreiros}}, \bibinfo {author} {\bibfnamefont {A.~A.}\ \bibnamefont
				{Zyuzin}}, \ and\ \bibinfo {author} {\bibfnamefont {J.~H.}\ \bibnamefont
				{Bardarson}},\ }\href {\doibase 10.1103/PhysRevB.96.115202} {\bibfield
			{journal} {\bibinfo  {journal} {Phys. Rev. B}\ }\textbf {\bibinfo {volume}
				{96}},\ \bibinfo {pages} {115202} (\bibinfo {year} {2017})}\BibitemShut
		{NoStop}%
		\bibitem [{\citenamefont {Chan}\ \emph {et~al.}(2017)\citenamefont {Chan},
			\citenamefont {Lindner}, \citenamefont {Refael},\ and\ \citenamefont
			{Lee}}]{Chan2017Jan}%
		\BibitemOpen
		\bibfield  {author} {\bibinfo {author} {\bibfnamefont {C.-K.}\ \bibnamefont
				{Chan}}, \bibinfo {author} {\bibfnamefont {N.~H.}\ \bibnamefont {Lindner}},
			\bibinfo {author} {\bibfnamefont {G.}~\bibnamefont {Refael}}, \ and\ \bibinfo
			{author} {\bibfnamefont {P.~A.}\ \bibnamefont {Lee}},\ }\href {\doibase
			10.1103/PhysRevB.95.041104} {\bibfield  {journal} {\bibinfo  {journal} {Phys.
					Rev. B}\ }\textbf {\bibinfo {volume} {95}},\ \bibinfo {pages} {041104}
			(\bibinfo {year} {2017})}\BibitemShut {NoStop}%
		\bibitem [{\citenamefont {K{\ifmmode\ddot{o}\else\"{o}\fi}nig}\ \emph
			{et~al.}(2017)\citenamefont {K{\ifmmode\ddot{o}\else\"{o}\fi}nig},
			\citenamefont {Xie}, \citenamefont {Pesin},\ and\ \citenamefont
			{Levchenko}}]{Konig2017Aug}%
		\BibitemOpen
		\bibfield  {author} {\bibinfo {author} {\bibfnamefont {E.~J.}\ \bibnamefont
				{K{\ifmmode\ddot{o}\else\"{o}\fi}nig}}, \bibinfo {author} {\bibfnamefont
				{H.-Y.}\ \bibnamefont {Xie}}, \bibinfo {author} {\bibfnamefont {D.~A.}\
				\bibnamefont {Pesin}}, \ and\ \bibinfo {author} {\bibfnamefont
				{A.}~\bibnamefont {Levchenko}},\ }\href {\doibase 10.1103/PhysRevB.96.075123}
		{\bibfield  {journal} {\bibinfo  {journal} {Phys. Rev. B}\ }\textbf {\bibinfo
				{volume} {96}},\ \bibinfo {pages} {075123} (\bibinfo {year}
			{2017})}\BibitemShut {NoStop}%
		\bibitem [{\citenamefont {Ma}\ \emph {et~al.}(2017)\citenamefont {Ma},
			\citenamefont {Xu}, \citenamefont {Chan}, \citenamefont {Zhang},
			\citenamefont {Chang}, \citenamefont {Lin}, \citenamefont {Xie},
			\citenamefont {Palacios}, \citenamefont {Lin}, \citenamefont {Jia},
			\citenamefont {Lee}, \citenamefont {Jarillo-Herrero},\ and\ \citenamefont
			{Gedik}}]{Ma2017Sep}%
		\BibitemOpen
		\bibfield  {author} {\bibinfo {author} {\bibfnamefont {Q.}~\bibnamefont
				{Ma}}, \bibinfo {author} {\bibfnamefont {S.-Y.}\ \bibnamefont {Xu}}, \bibinfo
			{author} {\bibfnamefont {C.-K.}\ \bibnamefont {Chan}}, \bibinfo {author}
			{\bibfnamefont {C.-L.}\ \bibnamefont {Zhang}}, \bibinfo {author}
			{\bibfnamefont {G.}~\bibnamefont {Chang}}, \bibinfo {author} {\bibfnamefont
				{Y.}~\bibnamefont {Lin}}, \bibinfo {author} {\bibfnamefont {W.}~\bibnamefont
				{Xie}}, \bibinfo {author} {\bibfnamefont {T.}~\bibnamefont {Palacios}},
			\bibinfo {author} {\bibfnamefont {H.}~\bibnamefont {Lin}}, \bibinfo {author}
			{\bibfnamefont {S.}~\bibnamefont {Jia}}, \bibinfo {author} {\bibfnamefont
				{P.~A.}\ \bibnamefont {Lee}}, \bibinfo {author} {\bibfnamefont
				{P.}~\bibnamefont {Jarillo-Herrero}}, \ and\ \bibinfo {author} {\bibfnamefont
				{N.}~\bibnamefont {Gedik}},\ }\href {\doibase 10.1038/nphys4146} {\bibfield
			{journal} {\bibinfo  {journal} {Nat. Phys.}\ }\textbf {\bibinfo {volume}
				{13}},\ \bibinfo {pages} {842} (\bibinfo {year} {2017})}\BibitemShut
		{NoStop}%
		\bibitem [{\citenamefont {Golub}\ and\ \citenamefont
			{Ivchenko}(2018)}]{Golub2018Aug}%
		\BibitemOpen
		\bibfield  {author} {\bibinfo {author} {\bibfnamefont {L.~E.}\ \bibnamefont
				{Golub}}\ and\ \bibinfo {author} {\bibfnamefont {E.~L.}\ \bibnamefont
				{Ivchenko}},\ }\href {\doibase 10.1103/PhysRevB.98.075305} {\bibfield
			{journal} {\bibinfo  {journal} {Phys. Rev. B}\ }\textbf {\bibinfo {volume}
				{98}},\ \bibinfo {pages} {075305} (\bibinfo {year} {2018})}\BibitemShut
		{NoStop}%
		\bibitem [{\citenamefont {Steiner}\ \emph {et~al.}(2017)\citenamefont
			{Steiner}, \citenamefont {Andreev},\ and\ \citenamefont
			{Pesin}}]{Steiner2017Jul}%
		\BibitemOpen
		\bibfield  {author} {\bibinfo {author} {\bibfnamefont {J.~F.}\ \bibnamefont
				{Steiner}}, \bibinfo {author} {\bibfnamefont {A.~V.}\ \bibnamefont
				{Andreev}}, \ and\ \bibinfo {author} {\bibfnamefont {D.~A.}\ \bibnamefont
				{Pesin}},\ }\href {\doibase 10.1103/PhysRevLett.119.036601} {\bibfield
			{journal} {\bibinfo  {journal} {Phys. Rev. Lett.}\ }\textbf {\bibinfo
				{volume} {119}},\ \bibinfo {pages} {036601} (\bibinfo {year}
			{2017})}\BibitemShut {NoStop}%
		\bibitem [{\citenamefont {Zyuzin}\ and\ \citenamefont
			{Zyuzin}(2018)}]{Zyuzin2018Jan}%
		\BibitemOpen
		\bibfield  {author} {\bibinfo {author} {\bibfnamefont {A.~A.}\ \bibnamefont
				{Zyuzin}}\ and\ \bibinfo {author} {\bibfnamefont {A.~{\relax Yu}.}\
				\bibnamefont {Zyuzin}},\ }\href {\doibase 10.1103/PhysRevB.97.041203}
		{\bibfield  {journal} {\bibinfo  {journal} {Phys. Rev. B}\ }\textbf {\bibinfo
				{volume} {97}},\ \bibinfo {pages} {041203} (\bibinfo {year}
			{2018})}\BibitemShut {NoStop}%
		\bibitem [{\citenamefont {Nguyen}\ and\ \citenamefont
			{Charlier}(2018)}]{Nguyen2018Jun}%
		\BibitemOpen
		\bibfield  {author} {\bibinfo {author} {\bibfnamefont {V.~H.}\ \bibnamefont
				{Nguyen}}\ and\ \bibinfo {author} {\bibfnamefont {J.-C.}\ \bibnamefont
				{Charlier}},\ }\href {\doibase 10.1103/PhysRevB.97.235113} {\bibfield
			{journal} {\bibinfo  {journal} {Phys. Rev. B}\ }\textbf {\bibinfo {volume}
				{97}},\ \bibinfo {pages} {235113} (\bibinfo {year} {2018})}\BibitemShut
		{NoStop}%
		\bibitem [{\citenamefont {Jia}\ \emph {et~al.}(2019)\citenamefont {Jia},
			\citenamefont {Zhang}, \citenamefont {Gao}, \citenamefont {Guo},
			\citenamefont {Yang}, \citenamefont {Hu}, \citenamefont {Bi}, \citenamefont
			{Xiang}, \citenamefont {Liu},\ and\ \citenamefont {Zhang}}]{Jia2019Jan}%
		\BibitemOpen
		\bibfield  {author} {\bibinfo {author} {\bibfnamefont {H.}~\bibnamefont
				{Jia}}, \bibinfo {author} {\bibfnamefont {R.}~\bibnamefont {Zhang}}, \bibinfo
			{author} {\bibfnamefont {W.}~\bibnamefont {Gao}}, \bibinfo {author}
			{\bibfnamefont {Q.}~\bibnamefont {Guo}}, \bibinfo {author} {\bibfnamefont
				{B.}~\bibnamefont {Yang}}, \bibinfo {author} {\bibfnamefont {J.}~\bibnamefont
				{Hu}}, \bibinfo {author} {\bibfnamefont {Y.}~\bibnamefont {Bi}}, \bibinfo
			{author} {\bibfnamefont {Y.}~\bibnamefont {Xiang}}, \bibinfo {author}
			{\bibfnamefont {C.}~\bibnamefont {Liu}}, \ and\ \bibinfo {author}
			{\bibfnamefont {S.}~\bibnamefont {Zhang}},\ }\href
		{https://www.science.org/doi/10.1126/science.aau7707} {\bibfield  {journal}
			{\bibinfo  {journal} {Science}\ }\textbf {\bibinfo {volume} {363}},\ \bibinfo
			{pages} {148} (\bibinfo {year} {2019})}\BibitemShut {NoStop}%
		\bibitem [{\citenamefont {Trescher}\ \emph {et~al.}(2015)\citenamefont
			{Trescher}, \citenamefont {Sbierski}, \citenamefont {Brouwer},\ and\
			\citenamefont {Bergholtz}}]{Trescher2015Mar}%
		\BibitemOpen
		\bibfield  {author} {\bibinfo {author} {\bibfnamefont {M.}~\bibnamefont
				{Trescher}}, \bibinfo {author} {\bibfnamefont {B.}~\bibnamefont {Sbierski}},
			\bibinfo {author} {\bibfnamefont {P.~W.}\ \bibnamefont {Brouwer}}, \ and\
			\bibinfo {author} {\bibfnamefont {E.~J.}\ \bibnamefont {Bergholtz}},\ }\href
		{\doibase 10.1103/PhysRevB.91.115135} {\bibfield  {journal} {\bibinfo
				{journal} {Phys. Rev. B}\ }\textbf {\bibinfo {volume} {91}},\ \bibinfo
			{pages} {115135} (\bibinfo {year} {2015})}\BibitemShut {NoStop}%
		\bibitem [{\citenamefont {Zyuzin}(2017)}]{Zyuzin2017}%
		\BibitemOpen
		\bibfield  {author} {\bibinfo {author} {\bibfnamefont {V.~A.}\ \bibnamefont
				{Zyuzin}},\ }\href {\doibase 10.1103/PhysRevB.95.245128} {\bibfield
			{journal} {\bibinfo  {journal} {Phys. Rev. B}\ }\textbf {\bibinfo {volume}
				{95}},\ \bibinfo {pages} {245128} (\bibinfo {year} {2017})}\BibitemShut
		{NoStop}%
		\bibitem [{\citenamefont {Sharma}\ \emph {et~al.}(2017)\citenamefont {Sharma},
			\citenamefont {Goswami},\ and\ \citenamefont {Tewari}}]{Sharma2017}%
		\BibitemOpen
		\bibfield  {author} {\bibinfo {author} {\bibfnamefont {G.}~\bibnamefont
				{Sharma}}, \bibinfo {author} {\bibfnamefont {P.}~\bibnamefont {Goswami}}, \
			and\ \bibinfo {author} {\bibfnamefont {S.}~\bibnamefont {Tewari}},\ }\href
		{\doibase 10.1103/PhysRevB.96.045112} {\bibfield  {journal} {\bibinfo
				{journal} {Phys. Rev. B}\ }\textbf {\bibinfo {volume} {96}},\ \bibinfo
			{pages} {045112} (\bibinfo {year} {2017})}\BibitemShut {NoStop}%
		\bibitem [{\citenamefont {Das}\ and\ \citenamefont
			{Agarwal}(2019{\natexlab{a}})}]{Das2019}%
		\BibitemOpen
		\bibfield  {author} {\bibinfo {author} {\bibfnamefont {K.}~\bibnamefont
				{Das}}\ and\ \bibinfo {author} {\bibfnamefont {A.}~\bibnamefont {Agarwal}},\
		}\href {\doibase 10.1103/PhysRevB.99.085405} {\bibfield  {journal} {\bibinfo
				{journal} {Phys. Rev. B}\ }\textbf {\bibinfo {volume} {99}},\ \bibinfo
			{pages} {085405} (\bibinfo {year} {2019}{\natexlab{a}})}\BibitemShut
		{NoStop}%
		\bibitem [{\citenamefont {Ma}\ \emph {et~al.}(2019)\citenamefont {Ma},
			\citenamefont {Jiang}, \citenamefont {Liu},\ and\ \citenamefont
			{Xie}}]{Ma2019a}%
		\BibitemOpen
		\bibfield  {author} {\bibinfo {author} {\bibfnamefont {D.}~\bibnamefont
				{Ma}}, \bibinfo {author} {\bibfnamefont {H.}~\bibnamefont {Jiang}}, \bibinfo
			{author} {\bibfnamefont {H.}~\bibnamefont {Liu}}, \ and\ \bibinfo {author}
			{\bibfnamefont {X.~C.}\ \bibnamefont {Xie}},\ }\href {\doibase
			10.1103/PhysRevB.99.115121} {\bibfield  {journal} {\bibinfo  {journal} {Phys.
					Rev. B}\ }\textbf {\bibinfo {volume} {99}},\ \bibinfo {pages} {115121}
			(\bibinfo {year} {2019})}\BibitemShut {NoStop}%
		\bibitem [{\citenamefont {Das}\ and\ \citenamefont
			{Agarwal}(2019{\natexlab{b}})}]{Das2019a}%
		\BibitemOpen
		\bibfield  {author} {\bibinfo {author} {\bibfnamefont {K.}~\bibnamefont
				{Das}}\ and\ \bibinfo {author} {\bibfnamefont {A.}~\bibnamefont {Agarwal}},\
		}\href {\doibase 10.1103/PhysRevB.100.085406} {\bibfield  {journal} {\bibinfo
				{journal} {Phys. Rev. B}\ }\textbf {\bibinfo {volume} {100}},\ \bibinfo
			{pages} {085406} (\bibinfo {year} {2019}{\natexlab{b}})}\BibitemShut
		{NoStop}%
		\bibitem [{\citenamefont {van~der Wurff}\ and\ \citenamefont
			{Stoof}(2019)}]{vanderWurff2019Jul}%
		\BibitemOpen
		\bibfield  {author} {\bibinfo {author} {\bibfnamefont {E.~C.~I.}\
				\bibnamefont {van~der Wurff}}\ and\ \bibinfo {author} {\bibfnamefont
				{H.~T.~C.}\ \bibnamefont {Stoof}},\ }\href {\doibase
			10.1103/PhysRevB.100.045114} {\bibfield  {journal} {\bibinfo  {journal}
				{Phys. Rev. B}\ }\textbf {\bibinfo {volume} {100}},\ \bibinfo {pages}
			{045114} (\bibinfo {year} {2019})}\BibitemShut {NoStop}%
		\bibitem [{\citenamefont {Ahn}\ \emph {et~al.}(2020)\citenamefont {Ahn},
			\citenamefont {Guo},\ and\ \citenamefont {Nagaosa}}]{Ahn2020Nov}%
		\BibitemOpen
		\bibfield  {author} {\bibinfo {author} {\bibfnamefont {J.}~\bibnamefont
				{Ahn}}, \bibinfo {author} {\bibfnamefont {G.-Y.}\ \bibnamefont {Guo}}, \ and\
			\bibinfo {author} {\bibfnamefont {N.}~\bibnamefont {Nagaosa}},\ }\href
		{\doibase 10.1103/PhysRevX.10.041041} {\bibfield  {journal} {\bibinfo
				{journal} {Phys. Rev. X}\ }\textbf {\bibinfo {volume} {10}},\ \bibinfo
			{pages} {041041} (\bibinfo {year} {2020})}\BibitemShut {NoStop}%
		\bibitem [{\citenamefont {Jiang}\ \emph {et~al.}(2021)\citenamefont {Jiang},
			\citenamefont {Wang}, \citenamefont {Bi}, \citenamefont {Fan}, \citenamefont
			{Zhao}, \citenamefont {Yu}, \citenamefont {Li},\ and\ \citenamefont
			{Wu}}]{Jiang2021Jun}%
		\BibitemOpen
		\bibfield  {author} {\bibinfo {author} {\bibfnamefont {B.}~\bibnamefont
				{Jiang}}, \bibinfo {author} {\bibfnamefont {L.}~\bibnamefont {Wang}},
			\bibinfo {author} {\bibfnamefont {R.}~\bibnamefont {Bi}}, \bibinfo {author}
			{\bibfnamefont {J.}~\bibnamefont {Fan}}, \bibinfo {author} {\bibfnamefont
				{J.}~\bibnamefont {Zhao}}, \bibinfo {author} {\bibfnamefont {D.}~\bibnamefont
				{Yu}}, \bibinfo {author} {\bibfnamefont {Z.}~\bibnamefont {Li}}, \ and\
			\bibinfo {author} {\bibfnamefont {X.}~\bibnamefont {Wu}},\ }\href {\doibase
			10.1103/PhysRevLett.126.236601} {\bibfield  {journal} {\bibinfo  {journal}
				{Phys. Rev. Lett.}\ }\textbf {\bibinfo {volume} {126}},\ \bibinfo {pages}
			{236601} (\bibinfo {year} {2021})}\BibitemShut {NoStop}%
		\bibitem [{\citenamefont {Wawrzik}\ \emph {et~al.}(2021)\citenamefont
			{Wawrzik}, \citenamefont {You}, \citenamefont {Facio}, \citenamefont {van~den
				Brink},\ and\ \citenamefont {Sodemann}}]{Wawrzik2021Jul}%
		\BibitemOpen
		\bibfield  {author} {\bibinfo {author} {\bibfnamefont {D.}~\bibnamefont
				{Wawrzik}}, \bibinfo {author} {\bibfnamefont {J.-S.}\ \bibnamefont {You}},
			\bibinfo {author} {\bibfnamefont {J.~I.}\ \bibnamefont {Facio}}, \bibinfo
			{author} {\bibfnamefont {J.}~\bibnamefont {van~den Brink}}, \ and\ \bibinfo
			{author} {\bibfnamefont {I.}~\bibnamefont {Sodemann}},\ }\href {\doibase
			10.1103/PhysRevLett.127.056601} {\bibfield  {journal} {\bibinfo  {journal}
				{Phys. Rev. Lett.}\ }\textbf {\bibinfo {volume} {127}},\ \bibinfo {pages}
			{056601} (\bibinfo {year} {2021})}\BibitemShut {NoStop}%
		\bibitem [{\citenamefont {Xiao}\ \emph {et~al.}(2006)\citenamefont {Xiao},
			\citenamefont {Yao}, \citenamefont {Fang},\ and\ \citenamefont
			{Niu}}]{Xiao2006}%
		\BibitemOpen
		\bibfield  {author} {\bibinfo {author} {\bibfnamefont {D.}~\bibnamefont
				{Xiao}}, \bibinfo {author} {\bibfnamefont {Y.}~\bibnamefont {Yao}}, \bibinfo
			{author} {\bibfnamefont {Z.}~\bibnamefont {Fang}}, \ and\ \bibinfo {author}
			{\bibfnamefont {Q.}~\bibnamefont {Niu}},\ }\href {\doibase
			10.1103/PhysRevLett.97.026603} {\bibfield  {journal} {\bibinfo  {journal}
				{Phys. Rev. Lett.}\ }\textbf {\bibinfo {volume} {97}},\ \bibinfo {pages}
			{026603} (\bibinfo {year} {2006})}\BibitemShut {NoStop}%
		\bibitem [{\citenamefont {Jia}\ \emph {et~al.}(2016)\citenamefont {Jia},
			\citenamefont {Li}, \citenamefont {Li}, \citenamefont {Shi}, \citenamefont
			{Liao}, \citenamefont {Yu},\ and\ \citenamefont {Wu}}]{Jia2016}%
		\BibitemOpen
		\bibfield  {author} {\bibinfo {author} {\bibfnamefont {Z.~Z.}\ \bibnamefont
				{Jia}}, \bibinfo {author} {\bibfnamefont {C.}~\bibnamefont {Li}}, \bibinfo
			{author} {\bibfnamefont {X.}~\bibnamefont {Li}}, \bibinfo {author}
			{\bibfnamefont {J.}~\bibnamefont {Shi}}, \bibinfo {author} {\bibfnamefont
				{Z.}~\bibnamefont {Liao}}, \bibinfo {author} {\bibfnamefont {D.}~\bibnamefont
				{Yu}}, \ and\ \bibinfo {author} {\bibfnamefont {X.~S.}\ \bibnamefont {Wu}},\
		}\href {\doibase 10.1038/ncomms13013} {\bibfield  {journal} {\bibinfo
				{journal} {Nat. Commun.}\ }\textbf {\bibinfo {volume} {7}},\ \bibinfo {pages}
			{13013} (\bibinfo {year} {2016})}\BibitemShut {NoStop}%
		\bibitem [{\citenamefont {Gooth}\ \emph {et~al.}(2017)\citenamefont {Gooth},
			\citenamefont {Niemann}, \citenamefont {Meng}, \citenamefont {Grushin},
			\citenamefont {Landsteiner}, \citenamefont {Gotsmann}, \citenamefont
			{Menges}, \citenamefont {Schmidt}, \citenamefont {Shekhar}, \citenamefont
			{Süß}, \citenamefont {Hühne}, \citenamefont {Rellinghaus}, \citenamefont
			{Felser}, \citenamefont {Yan},\ and\ \citenamefont {Nielsch}}]{Gooth2017}%
		\BibitemOpen
		\bibfield  {author} {\bibinfo {author} {\bibfnamefont {J.}~\bibnamefont
				{Gooth}}, \bibinfo {author} {\bibfnamefont {A.~C.}\ \bibnamefont {Niemann}},
			\bibinfo {author} {\bibfnamefont {T.}~\bibnamefont {Meng}}, \bibinfo {author}
			{\bibfnamefont {A.~G.}\ \bibnamefont {Grushin}}, \bibinfo {author}
			{\bibfnamefont {K.}~\bibnamefont {Landsteiner}}, \bibinfo {author}
			{\bibfnamefont {B.}~\bibnamefont {Gotsmann}}, \bibinfo {author}
			{\bibfnamefont {F.}~\bibnamefont {Menges}}, \bibinfo {author} {\bibfnamefont
				{M.}~\bibnamefont {Schmidt}}, \bibinfo {author} {\bibfnamefont
				{C.}~\bibnamefont {Shekhar}}, \bibinfo {author} {\bibfnamefont
				{V.}~\bibnamefont {Süß}}, \bibinfo {author} {\bibfnamefont
				{R.}~\bibnamefont {Hühne}}, \bibinfo {author} {\bibfnamefont
				{B.}~\bibnamefont {Rellinghaus}}, \bibinfo {author} {\bibfnamefont
				{C.}~\bibnamefont {Felser}}, \bibinfo {author} {\bibfnamefont
				{B.}~\bibnamefont {Yan}}, \ and\ \bibinfo {author} {\bibfnamefont
				{K.}~\bibnamefont {Nielsch}},\ }\href {\doibase 10.1038/nature23005}
		{\bibfield  {journal} {\bibinfo  {journal} {Nature}\ }\textbf {\bibinfo
				{volume} {547}},\ \bibinfo {pages} {324} (\bibinfo {year}
			{2017})}\BibitemShut {NoStop}%
		\bibitem [{\citenamefont {Zhang}\ \emph {et~al.}(2020)\citenamefont {Zhang},
			\citenamefont {Wang}, \citenamefont {Skinner}, \citenamefont {Bi},
			\citenamefont {Kozii}, \citenamefont {Cho}, \citenamefont {Zhong},
			\citenamefont {Schneeloch}, \citenamefont {Yu}, \citenamefont {Gu},
			\citenamefont {Fu}, \citenamefont {Wu},\ and\ \citenamefont
			{Zhang}}]{Zhang2020}%
		\BibitemOpen
		\bibfield  {author} {\bibinfo {author} {\bibfnamefont {W.}~\bibnamefont
				{Zhang}}, \bibinfo {author} {\bibfnamefont {P.}~\bibnamefont {Wang}},
			\bibinfo {author} {\bibfnamefont {B.}~\bibnamefont {Skinner}}, \bibinfo
			{author} {\bibfnamefont {R.}~\bibnamefont {Bi}}, \bibinfo {author}
			{\bibfnamefont {V.}~\bibnamefont {Kozii}}, \bibinfo {author} {\bibfnamefont
				{C.-W.}\ \bibnamefont {Cho}}, \bibinfo {author} {\bibfnamefont
				{R.}~\bibnamefont {Zhong}}, \bibinfo {author} {\bibfnamefont
				{J.}~\bibnamefont {Schneeloch}}, \bibinfo {author} {\bibfnamefont
				{D.}~\bibnamefont {Yu}}, \bibinfo {author} {\bibfnamefont {G.}~\bibnamefont
				{Gu}}, \bibinfo {author} {\bibfnamefont {L.}~\bibnamefont {Fu}}, \bibinfo
			{author} {\bibfnamefont {X.~S.}\ \bibnamefont {Wu}}, \ and\ \bibinfo {author}
			{\bibfnamefont {L.}~\bibnamefont {Zhang}},\ }\href {\doibase
			10.1038/s41467-020-14819-7} {\bibfield  {journal} {\bibinfo  {journal} {Nat.
					Commun.}\ }\textbf {\bibinfo {volume} {11}},\ \bibinfo {pages} {1046}
			(\bibinfo {year} {2020})}\BibitemShut {NoStop}%
		\bibitem [{\citenamefont {Liu}\ \emph {et~al.}(2018)\citenamefont {Liu},
			\citenamefont {Sun}, \citenamefont {Kumar}, \citenamefont {Muechler},
			\citenamefont {Sun}, \citenamefont {Jiao}, \citenamefont {Yang},
			\citenamefont {Liu}, \citenamefont {Liang}, \citenamefont {Xu}, \citenamefont
			{Kroder}, \citenamefont {Süß}, \citenamefont {Borrmann}, \citenamefont
			{Shekhar}, \citenamefont {Wang}, \citenamefont {Xi}, \citenamefont {Wang},
			\citenamefont {Schnelle}, \citenamefont {Wirth}, \citenamefont {Chen},
			\citenamefont {Goennenwein},\ and\ \citenamefont {Felser}}]{Liu2018}%
		\BibitemOpen
		\bibfield  {author} {\bibinfo {author} {\bibfnamefont {E.}~\bibnamefont
				{Liu}}, \bibinfo {author} {\bibfnamefont {Y.}~\bibnamefont {Sun}}, \bibinfo
			{author} {\bibfnamefont {N.}~\bibnamefont {Kumar}}, \bibinfo {author}
			{\bibfnamefont {L.}~\bibnamefont {Muechler}}, \bibinfo {author}
			{\bibfnamefont {A.}~\bibnamefont {Sun}}, \bibinfo {author} {\bibfnamefont
				{L.}~\bibnamefont {Jiao}}, \bibinfo {author} {\bibfnamefont {S.-Y.}\
				\bibnamefont {Yang}}, \bibinfo {author} {\bibfnamefont {D.}~\bibnamefont
				{Liu}}, \bibinfo {author} {\bibfnamefont {A.}~\bibnamefont {Liang}}, \bibinfo
			{author} {\bibfnamefont {Q.}~\bibnamefont {Xu}}, \bibinfo {author}
			{\bibfnamefont {J.}~\bibnamefont {Kroder}}, \bibinfo {author} {\bibfnamefont
				{V.}~\bibnamefont {Süß}}, \bibinfo {author} {\bibfnamefont
				{H.}~\bibnamefont {Borrmann}}, \bibinfo {author} {\bibfnamefont
				{C.}~\bibnamefont {Shekhar}}, \bibinfo {author} {\bibfnamefont
				{Z.}~\bibnamefont {Wang}}, \bibinfo {author} {\bibfnamefont {C.}~\bibnamefont
				{Xi}}, \bibinfo {author} {\bibfnamefont {W.}~\bibnamefont {Wang}}, \bibinfo
			{author} {\bibfnamefont {W.}~\bibnamefont {Schnelle}}, \bibinfo {author}
			{\bibfnamefont {S.}~\bibnamefont {Wirth}}, \bibinfo {author} {\bibfnamefont
				{Y.}~\bibnamefont {Chen}}, \bibinfo {author} {\bibfnamefont {S.~T.~B.}\
				\bibnamefont {Goennenwein}}, \ and\ \bibinfo {author} {\bibfnamefont
				{C.}~\bibnamefont {Felser}},\ }\href {\doibase 10.1038/s41567-018-0234-5}
		{\bibfield  {journal} {\bibinfo  {journal} {Nat. Phys.}\ }\textbf {\bibinfo
				{volume} {14}},\ \bibinfo {pages} {1125} (\bibinfo {year}
			{2018})}\BibitemShut {NoStop}%
		\bibitem [{\citenamefont {Liu}\ \emph {et~al.}(2019)\citenamefont {Liu},
			\citenamefont {Liang}, \citenamefont {Liu}, \citenamefont {Xu}, \citenamefont
			{Li}, \citenamefont {Chen}, \citenamefont {Pei}, \citenamefont {Shi},
			\citenamefont {Mo}, \citenamefont {Dudin}, \citenamefont {Kim}, \citenamefont
			{Cacho}, \citenamefont {Li}, \citenamefont {Sun}, \citenamefont {Yang},
			\citenamefont {Liu}, \citenamefont {Parkin}, \citenamefont {Felser},\ and\
			\citenamefont {Chen}}]{Liu2019}%
		\BibitemOpen
		\bibfield  {author} {\bibinfo {author} {\bibfnamefont {D.~F.}\ \bibnamefont
				{Liu}}, \bibinfo {author} {\bibfnamefont {A.~J.}\ \bibnamefont {Liang}},
			\bibinfo {author} {\bibfnamefont {E.~K.}\ \bibnamefont {Liu}}, \bibinfo
			{author} {\bibfnamefont {Q.~N.}\ \bibnamefont {Xu}}, \bibinfo {author}
			{\bibfnamefont {Y.~W.}\ \bibnamefont {Li}}, \bibinfo {author} {\bibfnamefont
				{C.}~\bibnamefont {Chen}}, \bibinfo {author} {\bibfnamefont {D.}~\bibnamefont
				{Pei}}, \bibinfo {author} {\bibfnamefont {W.~J.}\ \bibnamefont {Shi}},
			\bibinfo {author} {\bibfnamefont {S.~K.}\ \bibnamefont {Mo}}, \bibinfo
			{author} {\bibfnamefont {P.}~\bibnamefont {Dudin}}, \bibinfo {author}
			{\bibfnamefont {T.}~\bibnamefont {Kim}}, \bibinfo {author} {\bibfnamefont
				{C.}~\bibnamefont {Cacho}}, \bibinfo {author} {\bibfnamefont
				{G.}~\bibnamefont {Li}}, \bibinfo {author} {\bibfnamefont {Y.}~\bibnamefont
				{Sun}}, \bibinfo {author} {\bibfnamefont {L.~X.}\ \bibnamefont {Yang}},
			\bibinfo {author} {\bibfnamefont {Z.~K.}\ \bibnamefont {Liu}}, \bibinfo
			{author} {\bibfnamefont {S.~S.~P.}\ \bibnamefont {Parkin}}, \bibinfo {author}
			{\bibfnamefont {C.}~\bibnamefont {Felser}}, \ and\ \bibinfo {author}
			{\bibfnamefont {Y.~L.}\ \bibnamefont {Chen}},\ }\href
		{http://science.sciencemag.org/content/365/6459/1282.abstract} {\bibfield
			{journal} {\bibinfo  {journal} {Science}\ }\textbf {\bibinfo {volume}
				{365}},\ \bibinfo {pages} {1282} (\bibinfo {year} {2019})}\BibitemShut
		{NoStop}%
		\bibitem [{\citenamefont {Morali}\ \emph {et~al.}(2019)\citenamefont {Morali},
			\citenamefont {Batabyal}, \citenamefont {Nag}, \citenamefont {Liu},
			\citenamefont {Xu}, \citenamefont {Sun}, \citenamefont {Yan}, \citenamefont
			{Felser}, \citenamefont {Avraham},\ and\ \citenamefont
			{Beidenkopf}}]{Morali2019}%
		\BibitemOpen
		\bibfield  {author} {\bibinfo {author} {\bibfnamefont {N.}~\bibnamefont
				{Morali}}, \bibinfo {author} {\bibfnamefont {R.}~\bibnamefont {Batabyal}},
			\bibinfo {author} {\bibfnamefont {P.~K.}\ \bibnamefont {Nag}}, \bibinfo
			{author} {\bibfnamefont {E.}~\bibnamefont {Liu}}, \bibinfo {author}
			{\bibfnamefont {Q.}~\bibnamefont {Xu}}, \bibinfo {author} {\bibfnamefont
				{Y.}~\bibnamefont {Sun}}, \bibinfo {author} {\bibfnamefont {B.}~\bibnamefont
				{Yan}}, \bibinfo {author} {\bibfnamefont {C.}~\bibnamefont {Felser}},
			\bibinfo {author} {\bibfnamefont {N.}~\bibnamefont {Avraham}}, \ and\
			\bibinfo {author} {\bibfnamefont {H.}~\bibnamefont {Beidenkopf}},\ }\href
		{http://science.sciencemag.org/content/365/6459/1286.abstract} {\bibfield
			{journal} {\bibinfo  {journal} {Science}\ }\textbf {\bibinfo {volume}
				{365}},\ \bibinfo {pages} {1286} (\bibinfo {year} {2019})}\BibitemShut
		{NoStop}%
		\bibitem [{\citenamefont {Wang}\ \emph {et~al.}(2018)\citenamefont {Wang},
			\citenamefont {Xu}, \citenamefont {Lou}, \citenamefont {Liu}, \citenamefont
			{Li}, \citenamefont {Huang}, \citenamefont {Shen}, \citenamefont {Weng},
			\citenamefont {Wang},\ and\ \citenamefont {Lei}}]{Wang2018}%
		\BibitemOpen
		\bibfield  {author} {\bibinfo {author} {\bibfnamefont {Q.}~\bibnamefont
				{Wang}}, \bibinfo {author} {\bibfnamefont {Y.}~\bibnamefont {Xu}}, \bibinfo
			{author} {\bibfnamefont {R.}~\bibnamefont {Lou}}, \bibinfo {author}
			{\bibfnamefont {Z.}~\bibnamefont {Liu}}, \bibinfo {author} {\bibfnamefont
				{M.}~\bibnamefont {Li}}, \bibinfo {author} {\bibfnamefont {Y.}~\bibnamefont
				{Huang}}, \bibinfo {author} {\bibfnamefont {D.}~\bibnamefont {Shen}},
			\bibinfo {author} {\bibfnamefont {H.}~\bibnamefont {Weng}}, \bibinfo {author}
			{\bibfnamefont {S.}~\bibnamefont {Wang}}, \ and\ \bibinfo {author}
			{\bibfnamefont {H.}~\bibnamefont {Lei}},\ }\href {\doibase
			10.1038/s41467-018-06088-2} {\bibfield  {journal} {\bibinfo  {journal} {Nat.
					Commun.}\ }\textbf {\bibinfo {volume} {9}},\ \bibinfo {pages} {3681}
			(\bibinfo {year} {2018})}\BibitemShut {NoStop}%
		\bibitem [{\citenamefont {Guin}\ \emph {et~al.}(2019)\citenamefont {Guin},
			\citenamefont {Vir}, \citenamefont {Zhang}, \citenamefont {Kumar},
			\citenamefont {Watzman}, \citenamefont {Fu}, \citenamefont {Liu},
			\citenamefont {Manna}, \citenamefont {Schnelle}, \citenamefont {Gooth},
			\citenamefont {Shekhar}, \citenamefont {Sun},\ and\ \citenamefont
			{Felser}}]{Guin2019a}%
		\BibitemOpen
		\bibfield  {author} {\bibinfo {author} {\bibfnamefont {S.~N.}\ \bibnamefont
				{Guin}}, \bibinfo {author} {\bibfnamefont {P.}~\bibnamefont {Vir}}, \bibinfo
			{author} {\bibfnamefont {Y.}~\bibnamefont {Zhang}}, \bibinfo {author}
			{\bibfnamefont {N.}~\bibnamefont {Kumar}}, \bibinfo {author} {\bibfnamefont
				{S.~J.}\ \bibnamefont {Watzman}}, \bibinfo {author} {\bibfnamefont
				{C.}~\bibnamefont {Fu}}, \bibinfo {author} {\bibfnamefont {E.}~\bibnamefont
				{Liu}}, \bibinfo {author} {\bibfnamefont {K.}~\bibnamefont {Manna}}, \bibinfo
			{author} {\bibfnamefont {W.}~\bibnamefont {Schnelle}}, \bibinfo {author}
			{\bibfnamefont {J.}~\bibnamefont {Gooth}}, \bibinfo {author} {\bibfnamefont
				{C.}~\bibnamefont {Shekhar}}, \bibinfo {author} {\bibfnamefont
				{Y.}~\bibnamefont {Sun}}, \ and\ \bibinfo {author} {\bibfnamefont
				{C.}~\bibnamefont {Felser}},\ }\href {\doibase 10.1002/adma.201806622}
		{\bibfield  {journal} {\bibinfo  {journal} {Adv. Mater.}\ }\textbf {\bibinfo
				{volume} {31}},\ \bibinfo {pages} {1806622} (\bibinfo {year}
			{2019})}\BibitemShut {NoStop}%
		\bibitem [{\citenamefont {Costache}\ \emph {et~al.}(2012)\citenamefont
			{Costache}, \citenamefont {Bridoux}, \citenamefont {Neumann},\ and\
			\citenamefont {Valenzuela}}]{Costache2012}%
		\BibitemOpen
		\bibfield  {author} {\bibinfo {author} {\bibfnamefont {M.~V.}\ \bibnamefont
				{Costache}}, \bibinfo {author} {\bibfnamefont {G.}~\bibnamefont {Bridoux}},
			\bibinfo {author} {\bibfnamefont {I.}~\bibnamefont {Neumann}}, \ and\
			\bibinfo {author} {\bibfnamefont {S.~O.}\ \bibnamefont {Valenzuela}},\ }\href
		{\doibase 10.1038/nmat3201} {\bibfield  {journal} {\bibinfo  {journal} {Nat.
					Mater.}\ }\textbf {\bibinfo {volume} {11}},\ \bibinfo {pages} {199} (\bibinfo
			{year} {2012})}\BibitemShut {NoStop}%
		\bibitem [{\citenamefont {Raquet}\ \emph {et~al.}(2002)\citenamefont {Raquet},
			\citenamefont {Viret}, \citenamefont {Sondergard}, \citenamefont {Cespedes},\
			and\ \citenamefont {Mamy}}]{Raquet2002}%
		\BibitemOpen
		\bibfield  {author} {\bibinfo {author} {\bibfnamefont {B.}~\bibnamefont
				{Raquet}}, \bibinfo {author} {\bibfnamefont {M.}~\bibnamefont {Viret}},
			\bibinfo {author} {\bibfnamefont {E.}~\bibnamefont {Sondergard}}, \bibinfo
			{author} {\bibfnamefont {O.}~\bibnamefont {Cespedes}}, \ and\ \bibinfo
			{author} {\bibfnamefont {R.}~\bibnamefont {Mamy}},\ }\href {\doibase
			10.1103/PhysRevB.66.024433} {\bibfield  {journal} {\bibinfo  {journal} {Phys.
					Rev. B}\ }\textbf {\bibinfo {volume} {66}},\ \bibinfo {pages} {024433}
			(\bibinfo {year} {2002})}\BibitemShut {NoStop}%
		\bibitem [{\citenamefont {Shen}\ \emph {et~al.}(2019)\citenamefont {Shen},
			\citenamefont {Zeng}, \citenamefont {Zhang}, \citenamefont {Tong},
			\citenamefont {Ling}, \citenamefont {Xi}, \citenamefont {Wang}, \citenamefont
			{Liu}, \citenamefont {Wang}, \citenamefont {Wu},\ and\ \citenamefont
			{Shen}}]{Shen2019}%
		\BibitemOpen
		\bibfield  {author} {\bibinfo {author} {\bibfnamefont {J.}~\bibnamefont
				{Shen}}, \bibinfo {author} {\bibfnamefont {Q.}~\bibnamefont {Zeng}}, \bibinfo
			{author} {\bibfnamefont {S.}~\bibnamefont {Zhang}}, \bibinfo {author}
			{\bibfnamefont {W.}~\bibnamefont {Tong}}, \bibinfo {author} {\bibfnamefont
				{L.}~\bibnamefont {Ling}}, \bibinfo {author} {\bibfnamefont {C.}~\bibnamefont
				{Xi}}, \bibinfo {author} {\bibfnamefont {Z.}~\bibnamefont {Wang}}, \bibinfo
			{author} {\bibfnamefont {E.}~\bibnamefont {Liu}}, \bibinfo {author}
			{\bibfnamefont {W.}~\bibnamefont {Wang}}, \bibinfo {author} {\bibfnamefont
				{G.}~\bibnamefont {Wu}}, \ and\ \bibinfo {author} {\bibfnamefont
				{B.}~\bibnamefont {Shen}},\ }\href {\doibase 10.1063/1.5125722} {\bibfield
			{journal} {\bibinfo  {journal} {Appl. Phys. Lett.}\ }\textbf {\bibinfo
				{volume} {115}},\ \bibinfo {pages} {212403} (\bibinfo {year}
			{2019})}\BibitemShut {NoStop}%
		\bibitem [{not()}]{noteTEP}%
		\BibitemOpen
		\href@noop {} {}\bibinfo {note} {See Supplemental Material for the detailed
			theoretical calculations, which includes Refs. [48-54].}\BibitemShut {Stop}%
		\bibitem [{\citenamefont {Xiao}\ \emph {et~al.}(2010)\citenamefont {Xiao},
			\citenamefont {Chang},\ and\ \citenamefont {Niu}}]{Xiao2010}%
		\BibitemOpen
		\bibfield  {author} {\bibinfo {author} {\bibfnamefont {D.}~\bibnamefont
				{Xiao}}, \bibinfo {author} {\bibfnamefont {M.-C.}\ \bibnamefont {Chang}}, \
			and\ \bibinfo {author} {\bibfnamefont {Q.}~\bibnamefont {Niu}},\ }\href
		{\doibase 10.1103/RevModPhys.82.1959} {\bibfield  {journal} {\bibinfo
				{journal} {Rev. Mod. Phys.}\ }\textbf {\bibinfo {volume} {82}},\ \bibinfo
			{pages} {1959} (\bibinfo {year} {2010})}\BibitemShut {NoStop}%
		\bibitem [{\citenamefont {Shen}(2017)}]{Shen17book}%
		\BibitemOpen
		\bibfield  {author} {\bibinfo {author} {\bibfnamefont {S.-Q.}\ \bibnamefont
				{Shen}},\ }\href@noop {} {\emph {\bibinfo {title} {Topological Insulators:
					Dirac Equation in Condensed Matter}}},\ Vol.\ \bibinfo {volume} {187}\
		(\bibinfo  {publisher} {Springer},\ \bibinfo {year} {2017})\BibitemShut
		{NoStop}%
		\bibitem [{\citenamefont {Dai}\ \emph {et~al.}(2017)\citenamefont {Dai},
			\citenamefont {Du},\ and\ \citenamefont {Lu}}]{Dai17prl}%
		\BibitemOpen
		\bibfield  {author} {\bibinfo {author} {\bibfnamefont {X.}~\bibnamefont
				{Dai}}, \bibinfo {author} {\bibfnamefont {Z.}~\bibnamefont {Du}}, \ and\
			\bibinfo {author} {\bibfnamefont {H.-Z.}\ \bibnamefont {Lu}},\ }\href@noop {}
		{\bibfield  {journal} {\bibinfo  {journal} {Phys. Rev. Lett.}\ }\textbf
			{\bibinfo {volume} {119}},\ \bibinfo {pages} {166601} (\bibinfo {year}
			{2017})}\BibitemShut {NoStop}%
		\bibitem [{\citenamefont {Haug}\ and\ \citenamefont
			{Jauho}(2008)}]{Haug2008quantumkinetics}%
		\BibitemOpen
		\bibfield  {author} {\bibinfo {author} {\bibfnamefont {H.}~\bibnamefont
				{Haug}}\ and\ \bibinfo {author} {\bibfnamefont {A.-P.}\ \bibnamefont
				{Jauho}},\ }\href@noop {} {\emph {\bibinfo {title} {Quantum kinetics in
					transport and optics of semiconductors}}},\ Vol.~\bibinfo {volume} {2}\
		(\bibinfo  {publisher} {Springer},\ \bibinfo {year} {2008})\BibitemShut
		{NoStop}%
		\bibitem [{\citenamefont {Yip}(2015)}]{Yip15arXiv}%
		\BibitemOpen
		\bibfield  {author} {\bibinfo {author} {\bibfnamefont {S.-K.}\ \bibnamefont
				{Yip}},\ }\href@noop {} {\bibfield  {journal} {\bibinfo  {journal} {arXiv
					preprint arXiv:1508.01010}\ } (\bibinfo {year} {2015})}\BibitemShut {NoStop}%
		\bibitem [{\citenamefont {Shen}(2012)}]{TISbook}%
		\BibitemOpen
		\bibfield  {author} {\bibinfo {author} {\bibfnamefont {S.-Q.}\ \bibnamefont
				{Shen}},\ }\href@noop {} {\emph {\bibinfo {title} {Topological Insulators}}}\
		(\bibinfo  {publisher} {Springer-Verlag, Berlin Heidelberg},\ \bibinfo {year}
		{2012})\BibitemShut {NoStop}%
		\bibitem [{\citenamefont {Soldatov}\ \emph {et~al.}(2014)\citenamefont
			{Soldatov}, \citenamefont {Panarina}, \citenamefont {Hess}, \citenamefont
			{Schultz},\ and\ \citenamefont
			{Sch{\ifmmode\ddot{a}\else\"{a}\fi}fer}}]{Soldatov2014}%
		\BibitemOpen
		\bibfield  {author} {\bibinfo {author} {\bibfnamefont {I.~V.}\ \bibnamefont
				{Soldatov}}, \bibinfo {author} {\bibfnamefont {N.}~\bibnamefont {Panarina}},
			\bibinfo {author} {\bibfnamefont {C.}~\bibnamefont {Hess}}, \bibinfo {author}
			{\bibfnamefont {L.}~\bibnamefont {Schultz}}, \ and\ \bibinfo {author}
			{\bibfnamefont {R.}~\bibnamefont {Sch{\ifmmode\ddot{a}\else\"{a}\fi}fer}},\
		}\href {\doibase 10.1103/PhysRevB.90.104423} {\bibfield  {journal} {\bibinfo
				{journal} {Phys. Rev. B}\ }\textbf {\bibinfo {volume} {90}},\ \bibinfo
			{pages} {104423} (\bibinfo {year} {2014})}\BibitemShut {NoStop}%
	\end{thebibliography}
	%

	
\end{document}